\documentclass{emulateapj}
\usepackage{appendix}
\usepackage{color}

\received{}
\accepted{}
\slugcomment{Accepted, 18 February, 2014}
\shorttitle{{\it NuSTAR} View of Mrk 231}
\shortauthors{Teng et al.}

\begin{document}

\title{{\it NuSTAR} Reveals an Intrinsically X-ray Weak Broad Absorption Line Quasar in the Ultraluminous Infrared Galaxy Markarian~231}

\author{Stacy H. Teng\altaffilmark{1, 22}, W.N. Brandt\altaffilmark{2}, F.A. Harrison\altaffilmark{3}, B. Luo\altaffilmark{2}, D.M. Alexander\altaffilmark{4}, F.E. Bauer\altaffilmark{5, 6}, S.E. Boggs\altaffilmark{7}, F.E. Christensen\altaffilmark{8}, A., Comastri\altaffilmark{9}, W.W. Craig\altaffilmark{10, 7}, A.C. Fabian\altaffilmark{11}, D. Farrah\altaffilmark{12}, F. Fiore\altaffilmark{13}, P. Gandhi\altaffilmark{4}, B.W. Grefenstette\altaffilmark{3}, C.J. Hailey\altaffilmark{14}, R.C. Hickox\altaffilmark{15}, K.K. Madsen\altaffilmark{3}, A.F. Ptak\altaffilmark{16}, J.R. Rigby\altaffilmark{1}, G. Risaliti\altaffilmark{17, 18}, C. Saez\altaffilmark{5}, D. Stern\altaffilmark{19}, S. Veilleux\altaffilmark{20, 21}, D.J. Walton\altaffilmark{3}, D.R. Wik\altaffilmark{16, 22}, \& W.W. Zhang\altaffilmark{16}}

\altaffiltext{1}{Observational Cosmology Laboratory, NASA Goddard Space Flight Center, Greenbelt, MD 20771, USA; stacy.h.teng@nasa.gov}
\altaffiltext{2}{Department of Astronomy \& Astrophysics, 525 Davey Lab, The Pennsylvania State University, University Park, PA 16802, USA}
\altaffiltext{3}{Cahill Center for Astronomy and Astrophysics, California Institute of Technology, Pasadena, CA 91125, USA}
\altaffiltext{4}{Department of Physics, Durham University, South Road, Durham DH1 3LE, UK}
\altaffiltext{5}{Pontificia Universidad Cat\'olica de Chile, Departamento de Astronom\'ia y Astrof\'isica, Casilla 306, Santiago 22, Chile}
\altaffiltext{6}{Space Science Institute, 4750 Walnut Street, Suite 205, Boulder, CO 80301, USA}
\altaffiltext{7}{Space Sciences Laboratory, University of California, Berkeley, CA, 94720, USA}
\altaffiltext{8}{DTU Space - National Space Institute, Technical University of Denmark, Elektrovej 327, 2800 Lyngby, Denmark}
\altaffiltext{9}{INAF -- Osservatorio Astronomico di Bologna, Via Ranzani 1, 40127 Bologna, Italy}
\altaffiltext{10}{Lawrence Livermore National Laboratory, Livermore, CA 94550, USA}
\altaffiltext{11}{Institute of Astronomy, Madingley Road, Cambridge, CB3 0HA, UK}
\altaffiltext{12}{Department of Physics, Virginia Tech, Blacksburg, VA 24061, USA}
\altaffiltext{13}{Osservatorio Astronomico di Roma, via Frascati 33, 00040 Monteporzio Catone, Italy}
\altaffiltext{14}{Columbia Astrophysics Laboratory, Columbia University, New York, NY 10027, USA}
\altaffiltext{15}{Department of Physics and Astronomy, Dartmouth College, 6127 Wilder Laboratory, Hanover, NH 03755, USA}
\altaffiltext{16}{X-ray Astrophysics Laboratory, NASA Goddard Space Flight Center, Greenbelt, MD 20771, USA}
\altaffiltext{17}{INAF -- Osservatorio Astrofisico di Arcetri, Largo, E. Fermi 5, I-50125 Firenze, Italy}
\altaffiltext{18}{Harvard-Smithsonian Center for Astrophysics, 60 Garden St., Cambridge, MA 02138, USA}
\altaffiltext{19}{Jet Propulsion Laboratory, California Institute of Technology, Pasadena, CA 91109, USA}
\altaffiltext{20}{Department of Astronomy, University of Maryland, College Park, MD 20742, USA}
\altaffiltext{21}{Joint Space-Science Institute, University of Maryland, College Park, MD 20742, USA}
\altaffiltext{22}{NASA Postdoctoral Program Fellow}

\begin{abstract}
We present high-energy (3--30~keV) {\it NuSTAR} observations of the nearest quasar, the ultraluminous infrared galaxy (ULIRG) Markarian~231 (Mrk 231), supplemented with new and simultaneous low-energy (0.5--8~keV) data from {\it Chandra}.  The source was detected, though at much fainter levels than previously reported, likely due to contamination in the large apertures of previous non-focusing hard X-ray telescopes.  
The full band (0.5--30 keV) X-ray spectrum suggests the active galactic nucleus (AGN) in Mrk~231 is absorbed by a patchy and Compton-thin (N$_{\rm H} \sim1.2^{+0.3}_{-0.3}\times10^{23}$~cm$^{-2}$) column.  The intrinsic X-ray luminosity (L$_{\rm 0.5-30~keV}\sim1.0\times10^{43}$~erg~s$^{-1}$) is extremely weak relative to the bolometric luminosity where the 2--10~keV to bolometric luminosity ratio is $\sim$0.03\% compared to the typical values of  2--15\%.  Additionally, Mrk~231 has a low X-ray-to-optical power law slope ($\alpha_{\rm OX}\sim-1.7$).  It is a local example of a low-ionization broad absorption line (LoBAL) quasar that is intrinsically X-ray weak.  The weak ionizing continuum may explain the lack of mid-infrared [O IV], [Ne V], and [Ne VI] fine-structure emission lines which are present in sources with otherwise similar AGN properties. We argue that the intrinsic X-ray weakness may be a result of the super-Eddington accretion occurring in the nucleus of this ULIRG, and may also be naturally related to the powerful wind event seen in Mrk 231, a merger remnant escaping from its dusty cocoon.

\end{abstract}


\section{Introduction}
\label{sec:intro}



The presence of outflows in luminous quasars and active galactic nuclei (AGN) is most evident in broad absorption line (BAL) systems \citep{lynds67}.  Approximately 20\% of quasars show BAL features at some level \citep[e.g.,][]{gibson09}, understood to be the result of an observational sightline through AGN-driven wind material \citep[e.g.,][]{weymann91, ogle99, dipompeo13}.  In a common model for BAL quasars \citep[e.g.,][]{murray95, proga00}, the wind is launched from the accretion disk approximately 0.01--0.1 pc from the black hole and radiatively driven by ultraviolet (UV) line pressure.  However, X-ray emission from typical quasars would over-ionize the gas, quenching the line-driving mechanism and thus preventing the appearance of BAL features.  Therefore, this so-called ``disk-wind'' model requires that BAL quasars have lower levels of X-ray emission, which could either be intrinsic or be due to shielding very close to the black hole, near the base of the wind.  Indeed, BAL systems are very often observed to be X-ray faint \citep[e.g.,][]{g06, gibson09, wu10}, though observations have yet to provide a clear picture of the origin of this X-ray faintness.  Absorption of a nominal X-ray continuum has been established in several cases \citep[e.g.,][]{g02, grupe03, shemmer05, giustini08}, but there may be diversity among the population.  Following the terminology of other studies of X-ray emission in BAL quasars \citep[e.g.,][and references therein]{luo13}, the term ``X-ray weakness or faintness'' refers to the observed X-ray emission being significantly lower than expected given the optical-UV emission.  This may be caused by a normal X-ray continuum modified by absorption from gas and dust.  The term ``intrinsic X-ray weakness'' refers to an innate property of the AGN, which is one possible cause of the observed X-ray faintness.  

In order to investigate whether intrinsic X-ray weakness or shielding gas is the most likely explanation for the observed X-ray faintness of BAL quasars, sensitive observations above 10~keV are highly desirable as they can distinguish between the presence of Compton-thick ($N_{\rm H} \geq 1.5 \times 10^{24}$~cm$^{-2}$) shielding material and a less-obscured, intrinsically X-ray weak nucleus.  The {\it Nuclear Spectroscopic Telescope Array} ({\it NuSTAR}; \citealp{nustar}) is the first focusing high-energy X-ray telescope in orbit, and provides unprecedented sensitivity and angular resolution for high-energy, or hard (3--79 keV) X-ray photons.  Recently, {\it NuSTAR} observed two of the optically brightest BAL quasars known, PG~1004+130 and PG~1700+518, with the goal of investigating the origin of their X-ray faintness \citep{luo13}.  Unfortunately, low count rates prevented a definitive answer, but they conclusively showed that any absorption present must be Compton-thick.  \citet{luo13} were also able to demonstrate that about 17--40\% of BAL quasars are intrinsically X-ray weak by stacking {\it Chandra} observations of $z\approx1.5-3$ BAL quasars in the Large Bright Quasar Survey.  

Confirmation of intrinsically X-ray weak AGN in some BAL quasars would have important consequences regarding the origin of BAL features.  Instead of invoking shielding gas with an arbitrarily thick column density, intrinsic X-ray weakness would favor a scenario in which the BAL features are coupled to abnormally faint coronal X-ray emission, perhaps due to some mechanism that (periodically) quenches the X-ray corona (see \S~4.2 of \citealp{luo13} for discussion).  To this end, we have obtained an extremely
sensitive, broadband (0.5--30~keV) X-ray spectrum of Markarian 231 (Mrk 231), the nearest BAL quasar, using {\it Chandra} and {\it NuSTAR}.  

This paper is organized as follows:  \S2 details the multi-wavelength properties of Mrk~231; \S3 presents the new X-ray observations obtained for this study; \S4 discusses the previously reported hard X-ray spectra of Mrk~231, which are inconsistent with our new data; \S5 presents our modeling of the new broadband X-ray spectrum, \S6 discusses our favored scenario, that Mrk~231 is indeed intrinsically X-ray weak; and \S7 summarizes our results.  Throughout this paper, we adopt $H_0$ = 71 km s$^{-1}$ Mpc$^{-1}$, $\Omega_M = 0.27$, and $\Lambda = 0.73$ \citep{cosmo}.  Luminosities taken from the literature have been recalculated for our assumed cosmology.  

\begin{figure*}[ht]
\centering
\includegraphics[width=5.5in, angle=270]{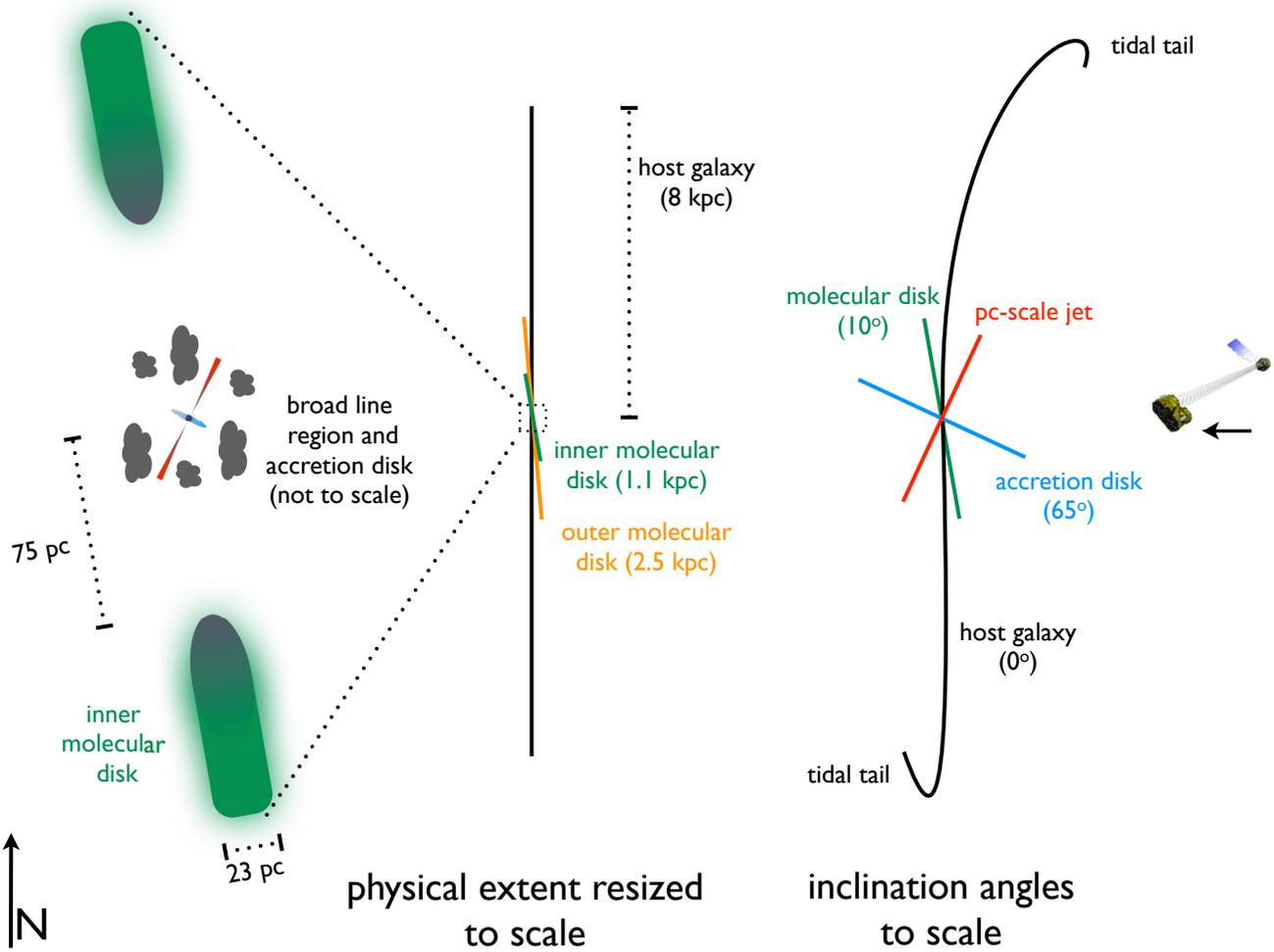}
\caption{Schematic showing the complexity of Mrk~231's structure and our assumed viewing geometry based on multiwavelength data.  The host galaxy is face-on on the plane of the sky with optical emission extending out to a radius of 8~kpc ($9\farcs8$) \citep[e.g.,][]{rv11}.  The thin (23~pc) molecular disk is tilted by 10$^\circ$ with the denser inner region extending from $\sim$75~pc ($0\farcs1$) to 1.1~kpc ($1\farcs3$) and the outer molecular disk extending out to 2.5~kpc ($3\farcs0$) \citep{downes98}.  The accretion disk is assumed to be orthogonal to the pc-scale jet, which is assumed to be inclined by 65$^{\circ}$ (see \S\ref{sec:torus}).  In this figure, north is up.}
\label{fig:geometry}
\end{figure*}

\section{The Multiwavelength Properties of Mrk~231}
\label{sec:multi}

One of the first BAL quasars observed by {\it NuSTAR} was Mrk~231, a galaxy with Type~1 optical spectral classification \citep[e.g.,][]{osterbrock78}.  It is the nearest quasar with a redshift of 0.042 \citep{boksenberg}.  At its distance of only 183~Mpc, 1$\arcsec$ subtends 0.82~kpc.  It is also one of the most infrared-luminous galaxies in the local universe \citep[L$_{\rm IR}\equiv{\rm L}_{8-1000~\mu{\rm m}}=3.6\times10^{12}$~L$_\odot$;][]{rbgs} and is therefore also considered an ultraluminous infrared galaxy (ULIRG).  Mrk~231 is a merger remnant, containing both an intense starburst and a luminous AGN.  The starburst has a star formation rate (SFR) of $\sim$140~M$_\odot$~yr$^{-1}$ \citep{rv13}, estimated from its total infrared luminosity.  However, there is considerable uncertainty in this value because of large scatters in the luminosity-to-SFR relations (see \S\ref{sec:simul} for details).  Modeling of the 1--1000~$\mu$m spectral energy distribution (SED) by \citet{farrah03} found an SFR of up to 450~M$_\odot$~yr$^{-1}$.  

Since the SED of ULIRGs is dominated by emission in the infrared band, the bolometric luminosity of ULIRGs is assumed to be 1.15 times the total infrared luminosity \citep[L$_{\rm bol}=1.15\times{\rm L_{IR}}$;][]{ks98}.  
The fractional contribution of the AGN and the starburst to the bolometric luminosity in composite sources like ULIRGs is very difficult to determine.  There is ambiguity in the literature concerning the exact AGN contribution (between 30--70\%) to the bolometric luminosity ($\sim$1.5$\times10^{46}$~erg~s$^{-1}$) for Mrk~231.  Modeling of the far-infrared SED by \citet{farrah03} implies $\sim$30\% contribution from the AGN.  More recently, \citet{vei09a} used six independent methods of measuring the AGN contribution to  the bolometric luminosity in their sample of ULIRGs and quasars.  These methods included fine structure line ratios, the EW of the 7.7~$\mu$m polycyclic aromatic hydrocarbon feature, and mid-infrared to far-infrared continuum flux ratios.  For Mrk~231, these methods found the AGN contributes 64 to 74\%, for an average of $\sim$70\%, to the bolometric luminosity.  For simplicity, we will refer to the AGN portion of the bolometric luminosity ($1.1\times10^{46}$~erg~s$^{-1}$) as L$_{\rm bol, AGN}$ assuming a 70\% AGN contribution to the bolometric luminosity.  


Mrk~231 is probably the best local example of quasar-mode feedback, where the active nucleus is thought to be driving a powerful kpc-scale outflow with velocities up to 1400~km~s$^{-1}$ \citep[e.g.,][]{rv13}.   This wide-angle wind is multi-phased, containing ionized \citep{rv13}, neutral \citep{rv13}, and molecular outflow phases \citep{feruglio, fischer, ga14}.  In the optical and UV, a BAL system is detected significantly in Mrk~231 in several transitions, including low-ionization species (e.g., Na~I~D, He~I, Ca~II, Mg~II, Mg~I, and Fe~II) and weakly in C~IV \citep[e.g.,][]{aw72, g02, g05, vei13a}.  Thus, Mrk~231 is categorized as a rare ($\sim$2\% of the BAL quasar population) iron low-ionization BAL (FeLoBAL) quasar \citep{dai12}.

The structure of Mrk~231 is complex; it contains spiral arms, multiple molecular disks, and jets in addition to the host galaxy and the AGN.  In Figure~\ref{fig:geometry}, we show a schematic of the structure of Mrk~231 and our assumed viewing geometry toward the nucleus described in this section.  Although considered to be a radio-quiet object, Mrk~231 has a weak radio jet that extends from the pc scale to the kpc scale \citep{carilli98, ulvestad}.  On the kpc scale, the jet becomes misaligned with the inner pc-scale jet \citep{carilli98, ulvestad}.  The kpc jet is thought to be providing additional acceleration to parts of the neutral outflow detected in spatially resolved maps in Na~I~D \citep{rv13}; single-dish observations in HI have also detected what is thought to be a weak jet-induced outflow  \citep{morganti11, teng13}.  CO observations of Mrk~231 by \citet{downes98} revealed two distinct molecular disks.  The denser inner molecular disk, extending from 75 to 1100 pc ( $\lesssim1\farcs3$), is nearly face on.  This molecular disk is likely the site of intense star formation (see \S~\ref{sec:var}).  Assuming an average dust temperature in the inner molecular disk of $\sim$200~K, heated by the AGN, the equivalent intrinsic column density derived from the \citet{downes98} measurements is $N_{\rm H} \sim 2.4\times10^{23}$~cm$^{-2}$.  Given the orientation of the pc-scale radio jet and assuming it is orthogonal to the accretion disk, the accretion disk has a different inclination than the inner molecular disk.  The outer molecular disk extends out to a radius of $\sim3\arcsec$, overlapping with the starburst arc observed to have multiple UV and optical star-forming knots \citep{surace98}.  

In the {\it Chandra} band, Mrk~231 is dominated by a point source, and has extended soft X-ray emission from the host galaxy \citep{g05}.  
The far-UV is largely unobscured and unpolarized, but the near-UV and optical spectra ($\sim$3000--4000~\AA) appear to be dust reddened by a screen of A$_{\rm V}\sim$7 mag \citep[e.g.,][]{vei13a}.  
Historic X-ray studies of Mrk~231 have found its spectrum to be relatively flat ($\Gamma\sim1.2$) suggestive of a reflection-dominated spectrum \citep{malreynolds, g02, g05, ptak03}.  However, the low Fe~K$\alpha$ equivalent width (EW) of $\sim$0.2~keV is difficult to reconcile with a  significant reflection component.  \citet{malreynolds} suggested that a large scattering fraction can dilute the Fe line EW.  \citet[][hereafter B04]{braito} and \citet[][hereafter P13]{piconcelli} have both claimed detections above 10~keV that imply Compton-thick absorption.  B04 measured an observed 0.5--10~keV luminosity of $\sim3\times10^{42}$~erg~s$^{-1}$; after correcting for absorption and removing the starburst component, the AGN luminosity in the 2--10~keV band is $\sim 5\times10^{43}$~erg~s$^{-1}$, about 0.5\% of the bolometric luminosity.  \citet{saez12} found no significant variations in the intrinsic AGN luminosity in the 0.5--8~keV band.  This is in agreement with P13 whose analysis suggests Compton-thick partial covering absorbers are responsible for the apparent flux variability below 10~keV seen on the time scale of years \citep[e.g..][]{g02, ptak03}.  Changes in the absorber are likely responsible for the apparent variability observed in Mrk~231.  These multiwavelength data imply complex geometries for the structure of Mrk~231. 


\begin{figure*}[ht]
\centering
\includegraphics[width=5.5in]{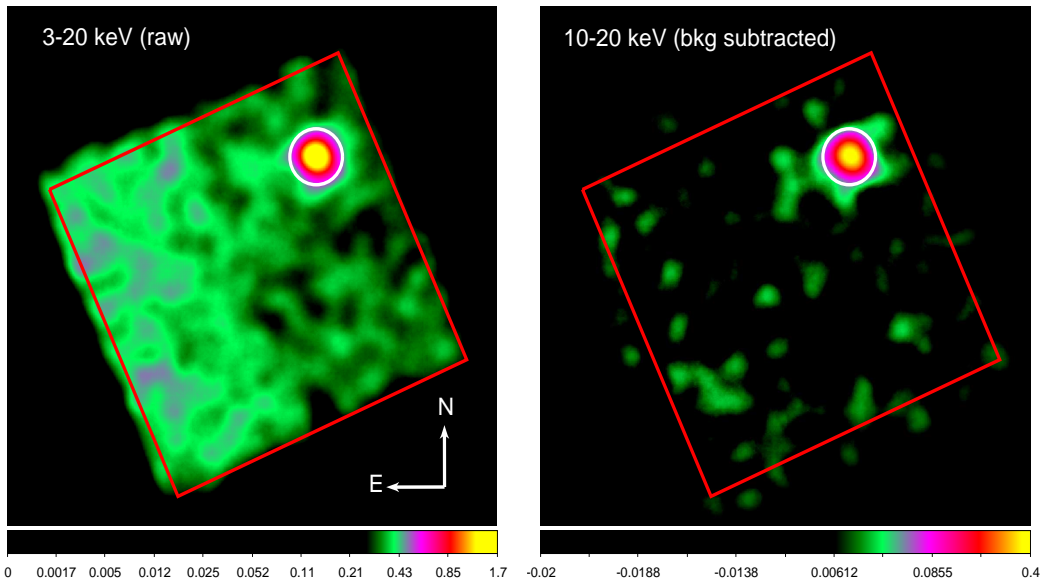}
\caption{The first epoch of {\it NuSTAR} FPMA observations of Mrk~231 smoothed with a sixteen-pixel radius Gaussian kernel.  Mrk~231 is clearly detected up to 20~keV.  We show the raw data in the 3--20~keV band on the left and data after the subtraction of the aperture background in the 10--20~keV band to demonstrate the quality of our background model.  In each panel, the white circle represents the 1$\arcmin$ radius source extraction region and the red square marks the $12\arcmin\times12\arcmin$ FOV of {\it NuSTAR}.  The color bar represents the total counts per pixel in log scale.}
\label{fig:extract}
\end{figure*}

\section{Observations and Data Reduction}
\label{sec:obs}

Mrk~231 was observed by {\em NuSTAR} for a total of 44~ks starting on UT 2012 August 27 (ObsID: 60002025002).  In order to check for hard X-ray variability, it was observed again for a total of 29~ks starting on UT 2013 May 9 (ObsID: 60002025004).  The first of these observations was scheduled to coincide with part of a 400~ks {\it Chandra} program with ACIS-S (PI: Veilleux; ObsID:13947).  

The {\em NuSTAR} data were reduced with the pipeline software NuSTARDAS version 0.11.1 and CALDB version 20130509 with the standard corrections applied.   After calibration and removal of bad time intervals, the final exposure is 41.1 and 28.6~ks on source for the first and second set of observations, respectively.  For the first epoch of {\it NuSTAR} data, the absolute astrometric offset relative to the Very Large Array position of Mrk~231 \citep{first} is $\sim$11$\arcsec$ and 10$\arcsec$ for focal plane module (FPM) A and B, respectively.  Similarly for the second epoch, the absolute astrometric offset is $\sim$8$\arcsec$ and 13$\arcsec$ for FPMA and B, respectively.  
The field of view (FOV) of each FPM is 12$\arcmin$ by 12$\arcmin$ \citep{nustar}.  On average, the first epoch of observations is off-axis by $\sim$4$\arcmin$.  The second epoch is off-axis by an average of $\sim$2$\arcmin$.  

The {\it Chandra} data were reduced using the standard ACIS-S scripts in CIAO version 4.5 and CALDB version 4.5.6.  The {\it Chandra} data reduction is fully described in Veilleux et al. (2014, in prep.).  The {\it Chandra} data contemporaneous with the first epoch of {\it NuSTAR} observations were extracted to complement our analysis of the hard X-ray data.  The strictly simultaneous good time intervals were extracted using the FTOOL ``mgtime''.  There is no obvious temporal variability detected during either the {\it NuSTAR} or {\it Chandra} exposure.  The two epochs of {\it NuSTAR} exposures also show no significant variability (see \S\ref{sec:nudata}).  

To be consistent between the two telescopes, source spectra were extracted using circles with 1$\arcmin$ radii for both the {\it NuSTAR} and {\it Chandra} data centered at the X-ray peak.  The {\it Chandra} data confirm that there are no contaminating point sources within the 1$\arcmin$ source extraction regions.  The {\it Chandra} background spectrum was extracted in the standard way, using a nearby circular extraction region where no obvious X-ray source is present.  The {\it NuSTAR} background extraction is more complicated due to the inhomogeneity of the background throughout each of the two detectors, caused by stray light incompletely blocked by the aperture stop (hereafter referred to as the aperture background).  
A simulated total background that combines both the aperture and instrumental backgrounds was produced, following the method of background simulation detailed in Wik et al. (2014, in prep.).  Briefly, the background models for the {\it NuSTAR} exposures were normalized to ``blank sky'' observations of the COSMOS and the Extended {\it Chandra} Deep Field-South observations accumulated by {\it NuSTAR} through May 2013.  For each epoch, a background spectrum was simulated from the model and then scaled to match the size of the source extraction region.  Conservatively, we estimate that the broadband systematic uncertainties in the derived background spectrum are $\lesssim$5\%.  In Figure~\ref{fig:extract}, we show our first epoch of {\it NuSTAR} data before and after subtraction of the aperture background which dominates at $\lesssim$20~keV. 

Spectral analysis was performed using HEASoft version 6.13.  The {\it Chandra} and {\it NuSTAR} source spectra were binned using the FTOOL ``grppha'' such that, in each bin, the total number of counts are four times the noise.  In the first epoch of {\it NuSTAR} observations, FPMA and FPMB detected 1012 and 1015 counts, respectively, from Mrk~231 in the 3--30~keV band.  Similarly, 819 and 771 counts were detected by FPMA and FPMB, respectively, in the second epoch.  When modeling the spectra, an additional constant factor, typically on the order of a few percent, is applied to account for the cross-normalization between FPMA and B and between FPMA and {\it Chandra} ACIS-S.  
We assumed the \citet{wilm} abundances and the \citet{vern} photoelectric cross sections in our spectral modeling with XSPEC.  The column density due to Galactic absorption is assumed to be N$_{\rm H, Gal} = 1.26\times10^{20}$~cm$^{-2}$ \citep{nh}. All errors quoted in this paper are at the 90\% confidence level ($\Delta\chi^2=2.706$ for a single parameter).

\section{Previous Hard X-ray Observations of Mrk~231}
\label{sec:contam}

Previous detections of Mrk~231 in the hard X-ray ($>10$~keV) have been claimed using the non-imaging {\it BeppoSAX} PDS (HPD $\sim1.4^\circ$; B04) and {\it Suzaku} PIN (HPD of 34$\arcmin$ by 34$\arcmin$ with a total square FOV $\sim$65\farcm5 on each side; P13).  The PDS and PIN 15--30~keV fluxes reported for Mrk~231 are 3.0 and $3.7\times10^{-12}$~erg~s$^{-1}$~cm$^{-2}$, respectively, in those publications.  B04 concluded through their modeling of the broadband 0.5--50~keV spectrum that Mrk~231 is a Compton-thick quasar with a column density of $\sim 1.8-2.6\times10^{24}$~cm$^{-2}$.  P13 supported this  conclusion using their more recent {\it Suzaku} observations, but a comparison of the B04 data with their own suggested Mrk~231 is variable below 10 keV.  They suggested that the variability is consistent with changes in the partial covering fraction of the Compton-thick absorber.

More recently, \citet{koss} reported a faint detection ($\sim4\sigma$) of Mrk~231 from the 70-month {\it Swift} BAT survey \citep{bat}.  The BAT has a nominal point spread function of $\sim$17$\arcmin$, smaller than both {\it Suzaku} and {\it BeppoSAX}.  By assuming a power law model ($\Gamma=1.9$), they estimate the 14--195~keV BAT flux of Mrk~231 to be $\sim5\times10^{-12}$~erg~s$^{-1}$~cm$^{-2}$, equivalent to $\sim1.2\times10^{-12}$~erg~s$^{-1}$~cm$^{-2}$ in the 15--30~keV band, approximately half of the values measured by B04 and P13.  For a random source to be considered a significant detection, the BAT team requires a flux limit of $4.8\sigma$ \citep{bat}.  This blind detection limit is above the peak BAT signal of Mrk~231.   Given the prior knowledge of the source position, the detection limit can be reduced to 3$\sigma$ because knowing the position of an astronomical source reduces the likelihood that the peak signal at that location is random noise \citep{koss}.  However, when the full BAT band image is divided into eight sub-band images, the position of the BAT peak changes from one band to another by several arcminutes, suggesting a likely spurious detection (W. Baumgartner, 2013, private communication).  The BAT flux of Mrk~231, which is at least a factor of two lower than the {\it BeppoSAX} and {\it Suzaku} detections, suggests that part of the hard X-ray flux measured by {\it BeppoSAX} and {\it Suzaku} has been resolved out by the smaller point spread function of BAT or is variable.  However, the BAT detection has significant uncertainties which can be addressed by the {\it NuSTAR} observations.

\begin{figure*}[ht]
\centering
\includegraphics[width=6in]{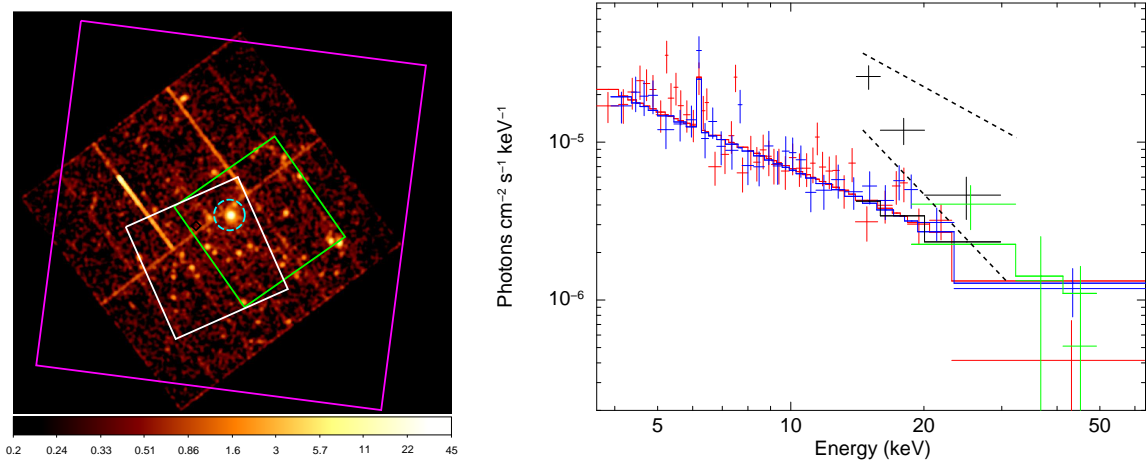}
\caption{{\it Left}: An image of the Mrk~231 field taken by {\it XMM-Newton}'s EPIC-pn in 0.5--10~keV (ObsID: 0081340201).  The data were minimally processed and smoothed with a 4-pixel radius Gaussian and clearly show the richness of the field surrounding Mrk~231.  The {\it NuSTAR} FOV of the first (white) and second (green) epochs are shown with the Mrk~231 spectral extraction region marked by the dashed cyan circle.  The {\it Suzaku} PIN HPD (34$\arcmin$ by 34$\arcmin$) is also shown in magenta.  The color bar represents log intensity in counts per pixel.  Multiple sources within the PIN FOV may have contaminated the PIN and {\it BeppoSAX} PDS (HPD$\sim$1.4$^\circ$) data.  {\it Right}: Unfolded combined {\em NuSTAR} FPMA/B (red/blue) spectra of Mrk~231 with PIN data from P13 (black), and PDS data from B04 (green).  The PIN data are nominally detected at only $\sim$5\% above the total background and the black dashed lines approximate the upper and lower limits of the PIN detection when the $\sim$3\% systematic error of the background is folded in.  The model shown is a simple power law best fit to the {\it NuSTAR} data.  The PIN spectrum (binned to 3$\sigma$) has a steeper slope than that found by {\em NuSTAR}.  However, the PDS data (also binned to 3$\sigma$) is not inconsistent with either the {\em NuSTAR} or PIN spectrum.}
\label{fig:contam}
\end{figure*}

\section{Modeling the X-ray Spectrum of Mrk 231}
\label{sec:spectra}

\subsection{The {\it NuSTAR} Data}
\label{sec:nudata}

The spectra from each epoch of {\it NuSTAR} observations are well-fit by a single power law model modified by only Galactic absorption and a narrow Gaussian to represent the Fe~K emission.  For the first epoch, the best-fit $\Gamma = 1.09^{+0.12}_{-0.11}$ ($\chi^2=75.9$ for 68 degrees of freedom).  The observed 15--30~keV flux is $1.5\pm0.2\times10^{-12}$~erg~s$^{-1}$~cm$^{-2}$, approximately the value measured by \citet{koss} despite their steeper assumed spectral slope.  The EW of the Fe~K emission line at $\sim$6.5~keV is 0.29$^{+0.12}_{-0.15}$~keV.  The second epoch spectrum has a similar best-fit $\Gamma$ of $1.33\pm0.14$ ($\chi^2=64.6$ for 63 degrees of freedom) with an observed 15--30~keV flux of $1.2\pm0.2\times10^{-12}$~erg~s$^{-1}$~cm$^{-2}$.  Fe~K emission is insignificant in the second epoch which has poorer photon statistics due to the shorter exposure.  Although the two epochs of {\it NuSTAR} data are separated by about seven months, there is no obvious sign of variability above 10~keV.  Therefore, to improve statistics, we created a combined spectrum of the two epochs of data using the ``addascaspec'' tool for each FPM.  The summed spectrum is used for the remainder of this paper, and is best-fit by a power law model with $\Gamma =1.22\pm0.08$ ($\chi^2=105.5$ for 108 degrees of freedom).  This model implies a 15--30~keV flux of $1.4\pm0.1\times10^{-12}$~erg~s$^{-1}$~cm$^{-2}$. 

The derived 15--30~keV observed flux from the {\it NuSTAR} data are approximately a third to a half of the values measured by B04 and P13; the {\it NuSTAR} spectrum is also significantly harder than those measured by B04 and P13.  Comparing the {\it NuSTAR} best-fit model to the {\it BeppoSAX} PDS and {\it Suzaku} PIN data, we find that the PDS and PIN detections are brighter and have steeper slopes (Figure~\ref{fig:contam}).  The PIN data are only detected at $\sim$5\% above the total background.  We were unable to check the accuracy of the ``tuned'' non-X-ray background provided by the {\it Suzaku} team since Earth occulted observations are unavailable for these data.  Taking into account the $\sim$3\% systematic error of the ``tuned'' background\footnote{http://heasarc.gsfc.nasa.gov/docs/suzaku/analysis/abc/node10.html\label{fn:pin}}, the lower limit of the PIN data shows an even steeper slope than the nominal value.  The change in spectral shape could be due to either contamination or variability.  The latter is unlikely because of the lack of variability in our {\it NuSTAR} epochs separated by seven months and the lack of intrinsic AGN variability measured in historic X-ray data \citep{g05, saez12}.  Unless the 10--30~keV X-ray spectrum of Mrk~231 hardened in the 14 months between the PIN and the first set of {\it NuSTAR} data, yet remained unchanged in the decade between the PDS and PIN observations, we conclude that the previous detections are contaminated by field sources outside of the {\it NuSTAR} FOV.  

Although not discussed in B04 or P13, contamination from nearby sources can be problematic for detectors with limited angular resolution such as the PDS or a large single-pixel FOV such as the PIN.   The PDS and PIN data match in flux and spectral index (Figure~\ref{fig:contam}), implying that if a contaminant exists for both sets of observations, then it must lie within the PIN FOV.  Taking advantage of the larger FOV of {\it XMM-Newton} EPIC-pn over that of {\it Chandra} ACIS-S, we show the 17~ks 0.5--10~keV image of the Mrk~231 field taken by {\it XMM-Newton} in 2001 (ObsID: 0081340201) in Figure~\ref{fig:contam}.  The field surrounding Mrk~231 is richly populated with point sources.  Mrk~231 dominates the field in the 0.5--10~keV band, so it is reasonable to assume that it also dominates the field above 10~keV as other authors have done.  It is unlikely that the source of contamination is a single object brighter than Mrk~231.  However, with so many field sources, some may even be outside of the 27$\arcmin$ by 26$\arcmin$ {\it XMM-Newton} FOV; it is likely that the integrated spectrum of multiple sources fainter than Mrk~231 add to the PIN and PDS detections.  A number of point sources detected by {\it XMM-Newton} within the {\it NuSTAR} FOVs are weakly detected by {\it NuSTAR} at 10--20~keV (see Figure~\ref{fig:extract}) and there may be brighter sources outside the {\it NuSTAR} FOV contributing to the PIN detection. 

While we cannot definitively rule out the possibility that the change in Mrk~231 flux above 10~keV is due to intrinsic variability, we conclude that the {\it BeppoSAX} and PIN detections were probably due to contamination for the following reasons.  

\begin{enumerate}
\item The PIN detection was at only 5.1\% above the background between 15--40~keV.  At the 90\% confidence level, the reproducibility of the total PIN background in this energy range is $\sim$3\%$^{23}$.  Therefore, about 5\% of the time, the background is higher than that modeled, so the PIN detection could be consistent with a non-detection.  Similarly, Mrk~231 was detected by {\it BeppoSAX} PDS at a rate of 0.066$\pm$0.022 counts per second \citep{braito}.  The full band systematic error for PDS is 0.020$\pm$0.015 counts per second\footnote{http://heasarc.gsfc.nasa.gov/docs/sax/abc/saxabc/saxabc.html\#SECTION00052300000000000000.  The typical systematic error in the PDS count rate is given by a technical report by Guainazzi \& Matteuzzi (1998) that is no longer available publicly.  The authors compared PDS spectra from a number of blank fields to derive the residual count rate.  Since the original document is no longer available online, the error is assumed to be at 90\% confidence level similar to the confidence quoted on other errors in the current document and is typical of X-ray studies.}.  Thus, the PDS detection is marginal and could also be consistent with a non-detection.   
\item Using the hard band (2.5--8~keV) $\log N - \log S$ \citep[e.g.,][]{lognlogs} derived from random fields from all over the sky and assuming a flux $S$ equal to the hard band flux of Mrk~231, there is a 33\% chance a background source is of that flux or brighter within one deg$^2$ (FOV of the PIN).  Since the PIN response falls off linearly with off-axis angle, to produce the assumed detected flux, it is most likely that the source has to be brighter than $S$.  Conservatively, within the central 0.25 deg$^2$, the chances of contamination by a source of flux $\geq 2S$ is better than 3\%.  Because the PIN's lack of imaging capability cannot rule out serendipitous sources, we expect the chance of a source falling within the PIN FOV of one deg$^2$ is better than 10\%.  This probability increases when we consider sources fainter than $2S$ contaminating the field and detectors with larger FOV such as that of {\it BeppoSAX}.  The 2.5--8~keV $\log N - \log S$ may not match perfectly with the 10--30~keV $\log N - \log S$, but the only $\log N - \log S$ currently available above 10~keV is from the BAT survey which only covers the brightest sources.  
\item As stated above, the {\it NuSTAR} 10--30~keV flux is consistent with the BAT results integrated over nearly five years \citep{koss}.  Since the temporal coverage is so long, we can assume that the BAT value is the typical average flux of Mrk 231.  The derived {\it BeppoSAX} and PIN fluxes are more than twice the BAT limit.  Therefore, it would be a remarkable coincidence if the AGN flared twice only at times observed by detectors with much larger PSFs than {\it NuSTAR}.  
\item If the AGN flared when observed by {\it BeppoSAX} and PIN but not when observed by {\it NuSTAR}, the lack of variability below 10~keV or change in the Fe line EW is puzzling \citep{g05, saez12}.  It would require a contrived scenario to have a Compton-thick cloud present to obscure the hard X-ray continuum only when the 10--30~keV flux is in the high state so that continuum variability is not detected below 10~keV or without change to the Fe line EW.  
\end{enumerate}
Therefore, we conclude that contamination is the simplest explanation given all the circumstances. 

\subsection{The Broadband {\it Chandra} and {\it NuSTAR} Spectrum}
\label{sec:simul}

\begin{figure*}[ht]
\centering
\includegraphics[width=6.5in]{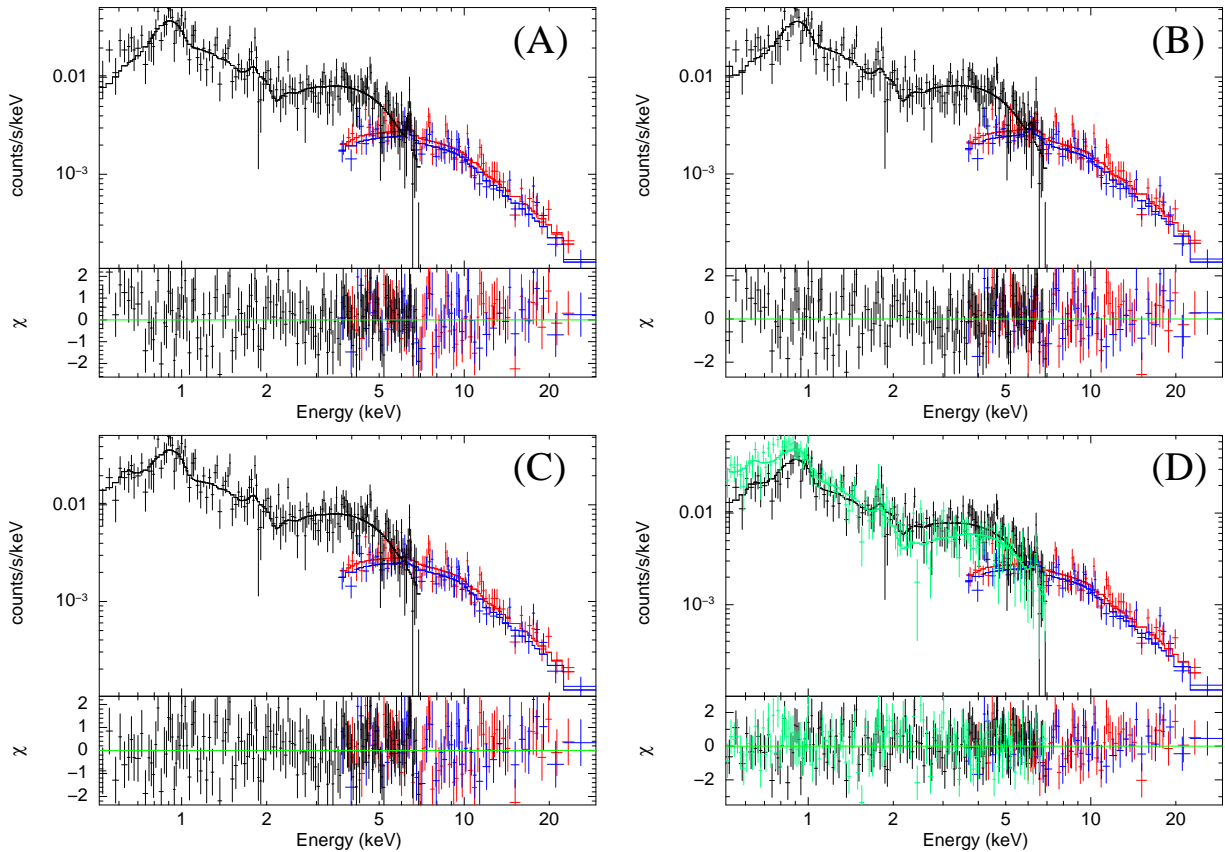}
\caption{{\it Chandra} (black) and {\it NuSTAR} FPMA/B (red/blue) spectra of Mrk~231 shown with various models.  (A):  Single power law modified by a single partial covering neutral absorber. (B): MYTorus without independent constraints on the starburst component.  (C):  MYTorus model with the X-ray emission from the star formation of the host galaxy constrained.  (D): The best-fit MYTorus model simultaneously fit to 2003 (green) and 2012 (black) {\it Chandra} data. All models result in a similar quality of fit, though as described in \S~\ref{sec:spectra}, several of the models are deemed improbable because of required galaxy-scale Compton-thick columns or atypical SFR-to-X-ray luminosity values. }  
\label{fig:fits}
\end{figure*}

Our {\it Chandra} and {\it NuSTAR} observations of Mrk~231 comprise the most sensitive broadband X-ray spectrum of this object to date.  In this section, we explore various models in order to constrain the three-dimensional geometry of the AGN and its associated absorber system as well as the intrinsic luminosity of the AGN.   As stated previously, the {\it NuSTAR}-only spectra are well-fit by a single power law model.  With the {\it Chandra} data anchoring the lower energies, we modeled the full 0.5--30~keV spectrum ($\chi^2=314.7$ for 261 degrees of freedom) with a MEKAL gas temperature of 0.82$\pm$0.04 keV and a very flat power law ($\Gamma = 1.20^{+0.09}_{-0.08}$) modified by a small column (N$_{\rm H} = 2.9^{+0.7}_{-0.5} \times10^{22}$~cm$^{-2}$).  There is no telltale sign of a Compton-reflection hump above 10~keV that would be indicative of significant reflection; this is  corroborated by the weak Fe~K line (EW$\sim$0.30~keV).  If Mrk~231 were to contain a normal AGN with a typical photon index \citep[e.g., $1.5<\Gamma<2.2$;][]{nandra, reeves}, then its apparent flat X-ray spectrum would require more complex models and geometries.   

Since Mrk~231 is unresolved by the {\it NuSTAR} beam, star formation is responsible for a non-negligible fraction of the total X-ray emission detected by {\it NuSTAR}.  For this reason, our modeling will include a starburst and an AGN component.  The starburst component consists of a MEKAL model representing the hot gas associated with the starburst and a power law model with $\Gamma = 1.1$ and cutoff energy of 10~keV representing the non-thermal component of high mass X-ray binaries \citep[HMXBs; e.g.,][; B04]{hmxb, persic02, hmxb2}.  The AGN component is represented by different versions of an absorbed power law model.  

In the following subsections, we present a detailed discussion of our spectral modeling of the broadband X-ray spectra.  We first applied simple geometry models such as those presented by other authors (e.g., B04, P13).  We then used the more sophisticated MYTorus model \citep{mytorus} to model the AGN emission.  In the process, we find that it is difficult to constrain the various starburst and AGN components independently; these models result in poor constraints on the HMXB component.  The derived HMXB luminosities from SFR-L$_{\rm X}$ scaling relations are too high given the infrared-derived SFR \citep{lehmer, mineo12b} due to degeneracies in the spectral models. Exploiting the spatial resolution of {\it Chandra}, we separated out the X-ray emission into nuclear and host galaxy components.  By fixing the host galaxy component, we are able to robustly measure the AGN emission.  We conclude that the AGN is extremely under luminous in the X-rays.  

\subsubsection{Simple Geometries}
\label{sec:simple}

First, we modeled the broadband spectrum from the 1$\arcmin$ aperture with the AGN component being a power law modified by a single partial covering absorber.  The AGN photon index ($\Gamma_{\rm AGN}$) tended toward a flatter slope than the canonical value of 1.8 ($\Gamma_{\rm AGN} \sim1.4$).  We tried fixing $\Gamma_{\rm AGN}$ to 1.8, but the resulting fit is a poor model of the spectrum which underestimates the flux above $\sim$10~keV.  Thus, we allowed $\Gamma_{\rm AGN}$ to be free.  The best-fit values are listed in Column 2 of Table~\ref{tab:fits} and the spectrum is shown in Figure~\ref{fig:fits}.  The derived gas temperature of $\sim$0.81~keV is consistent with those previously measured in ULIRGs \citep[e.g.,][]{ptak03, teng10}.  Assuming a total SFR of $\sim$140~M$_\odot$~yr$^{-1}$ derived from the total infrared luminosity of the non-AGN component, the expected X-ray luminosity of the starburst is derived using the SFR-L$_{\rm X}$ relations found by \citet{mineo12a, mineo12b}.  The luminosity we find for the hot gas component, $\sim1.5\times10^{41}$~erg~s$^{-1}$, is consistent with the expected value of $\sim7.3\times10^{40}$~erg~s$^{-1}$ to within the 0.34~dex scatter of the \citet{mineo12b} relation.  However, the 0.5--8~keV luminosity of the HMXB power law component is nearly a factor of 9 lower than the expected value of $\sim3.6\times10^{41}$~erg~s$^{-1}$ based on the \citet{mineo12a} relation.  We assumed no obscuration for this component, but an average column of $\sim3\times10^{24}$~cm$^{-2}$ is required in order for the derived HMXB luminosity to be consistent with prediction.  It is improbable that such a significant column is present over the whole galaxy.  The absorption-corrected 2--10~keV AGN luminosity only accounts for $\sim$0.05\% of L$_{\rm bol, AGN}$, far lower than the typical value of $\sim$2\% in luminous quasars \citep{elvis}.

Following \citet{g05} and P13, who modified the B04 model of Mrk~231 to explain the spectral variability at $<10$~keV, we applied a double partial covering absorber model to the new broadband data.   Such a model is a poor fit to the data.  Since there is no sign of Compton-reflection in the {\it NuSTAR} spectrum, the reflection component is statistically unnecessary ($\Delta \chi^2$ = 3.7 for 2 degrees of freedom).  The need for the reflection component by P13 to fit their {\it Suzaku} data is possibly due to the likely contamination discussed in \S\ref{sec:contam}.   Therefore, we do not discuss this model further. 


\subsubsection{Toroidal Geometries}
\label{sec:torus}

We also implemented a toroidal model that self-consistently accounts for the transmitted and scattered AGN emission in a torus geometry in the Compton-thick regime \citep[MYTorus;][]{mytorus}.  While the model assumes only neutral material, it is also relevant for partially ionized material since Compton scattering dominates in the Compton-thick regime.  The MYTorus model is a table model that contains three main components: the direct (or zeroth order) AGN component, the Compton-scattered component, and some fluorescence emission line components for self-consistent modeling.  An additional Compton-thin power law component can be added to represent ``leaked'' emission due to a partial covering absorber.  The MYTorus table models are relevant for $\Gamma$ in the range of 1.4--2.5 and N$_{\rm H}$ up to 10$^{25}$~cm$^{-2}$. 

The MYTorus model defaults to a half-opening angle of 60$^\circ$ which corresponds to a covering factor of 0.5.  Therefore, for an inclination angle of 60$^\circ$, the sightline to the center skirts the edge of the absorber.  In the modeling, the strength of the Fe~K line provides the main constraints on the inclination angle and the column density.  Since the Fe~K complex in Mrk~231 is so weak, we held the inclination angle fixed at 65$^\circ$ in order to derive a robust value for the column density.  The assumption of 65$^\circ$ is based on several observational factors including the position angle of the inner pc-scale jet \citep{ulvestad}, which does not contribute significantly to the X-ray emission \citep{g02}.  Additionally, the profile of the Ly$\alpha$ line suggests the accretion disk must have an inclined geometry \citep[inclination $\gtrsim$45$^\circ$;][]{vei13a}.  The partial covering result of P13 suggests that the line of sight through the torus must be near the ``fuzzy'' edge of the torus \citep{mytorus}.  The robustness of our results based on the assumed inclination of 65$^\circ$ is discussed below in \S\ref{sec:var}.


By allowing all the relevant model parameters to be free, the best-fit model requires a Compton-thin component, which is akin to the ``leaked'' direct emission in a partial covering absorber model favored by P13.  The best-fit model suggests that the AGN is border-line Compton-thick (N$_{\rm H} \sim 2.1^{+1.7}_{-1.7}\times10^{24}$~cm$^{-2}$).  However, the starburst element is poorly constrained.  The derived luminosity for the HMXB component is unphysical  (more than ten times too luminous) with respect to the SFR-to-X-ray luminosity scaling relations assuming an infrared-derived SFR of 140~M$_\odot$~yr$^{-1}$ \citep{lehmer, mineo12b}.

To better constrain the HMXB contribution to the broadband X-ray spectrum, we took advantage of {\it Chandra}'s excellent spatial resolution and excised the inner 1$\arcsec$ (0.82~kpc) emission from the spectrum.  Using this host-only {\it Chandra} spectrum, we measured the expected contribution of the starburst to the X-ray emission separate from that of the AGN.  Accounting for the wings of the ACIS point spread function ($\sim$20\% of the total counts lie beyond 1$\arcsec$), we find no significant contribution from HMXBs in the host spectrum (e.g., Veilleux et al. 2014, in prep.). The host-only {\it Chandra} spectrum is best-fit by two MEKAL components with temperatures at 0.25$^{+0.05}_{-0.06}$ and 0.93$^{+0.11}_{-0.10}$~keV.  
The requirement of a two-temperature MEKAL component is consistent with the two temperatures derived for the extended X-ray emission by \citet{g02}.  
The best-fit spectrum is shown in Figure~\ref{fig:fits}.  Contrary to the results from when the starburst component is allowed to vary, Mrk~231 is, in fact, Compton-thin (N$_{\rm H} \sim 1.3^{+0.03}_{-0.02}\times10^{23}$~cm$^{-2}$), but with $\Gamma_{\rm AGN}$ of 1.40--1.47.  Additionally, compared to L$_{\rm bol, AGN}$, the intrinsic 2--10~keV X-ray luminosity of Mrk~231 is only 0.04\% .  
The modeling of the X-ray emission from circumnuclear star formation is discussed in the following subsection.

\subsubsection{The Preferred Model and the Variability of Mrk~231}
\label{sec:var}

\begin{figure*}[ht]
\centering
\includegraphics[width=5in, angle=270]{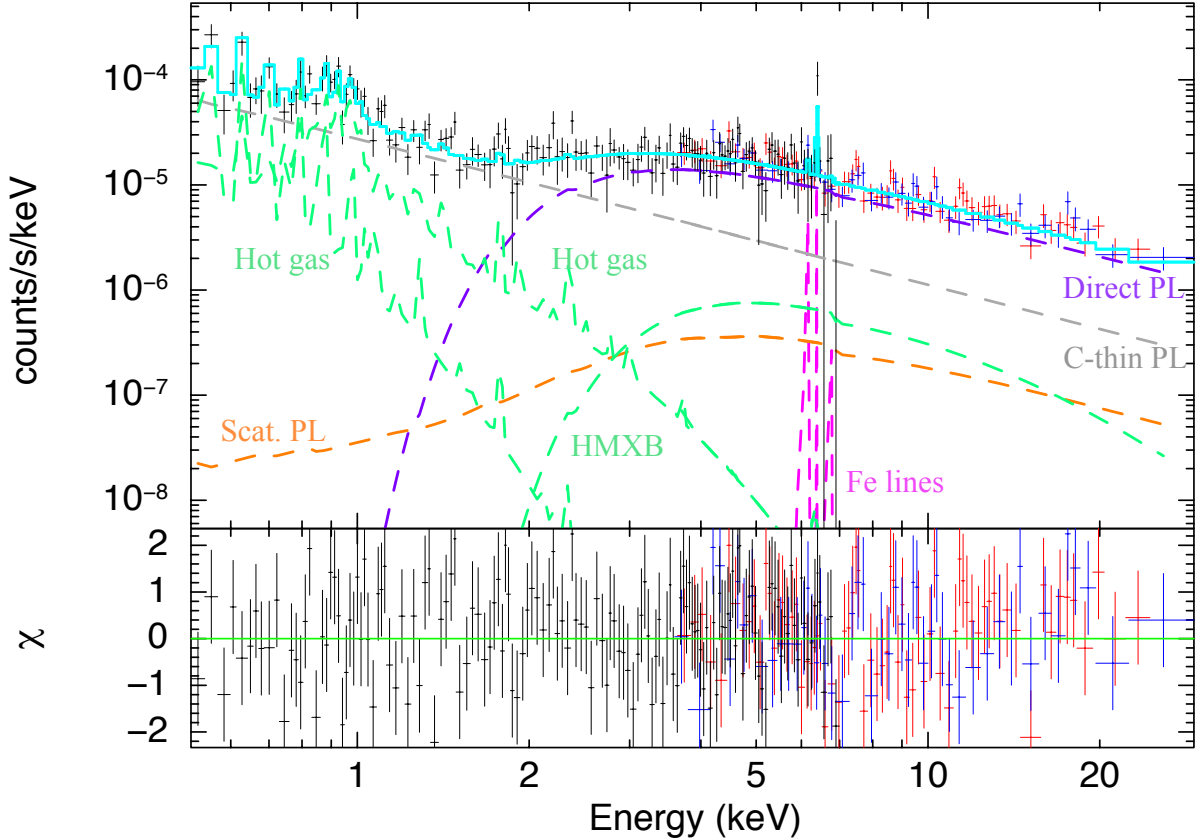}
\caption{Unfolded {\it Chandra} and {\it NuSTAR} spectra shown with the best-fit model components (dashed lines) with the total model in cyan.  The starburst MEKAL and HMXB components are in green.  The AGN component includes Compton-thin AGN emission unaffected by the absorber (grey), the ``zeroth'' order direct emission from the AGN transmitted through the absorber (purple), the scattered AGN emission through the absorber (orange), and the Fe~K emission lines modeled self-consistently within MYTorus as well as the 6.7~keV Fe line (magenta).}  
\label{fig:fitsmod}
\end{figure*}

From the previous sections, it is clear that accurate accounting of the starburst contribution to the X-ray luminosity is key to finding a robust measurement of the intrinsic AGN luminosity.  Due to the lack of spatial resolution and the effects of obscuration within 1$\arcsec$ of the nucleus, it is difficult to separate the starburst and the AGN emission in the X-rays.  Spatially resolved integral field unit maps of Mrk~231 in H$\alpha$ from \citet{rv13} place poor constraints on the SFR in the nuclear region because of heavy extinction.  Similar maps in Pa$\alpha$ taken recently with adaptive optics show that the majority of star formation is taking pace within the inner 0.5~kpc ($0\farcs6$) diameter of the nucleus (D. Rupke 2013, private communication).  

Consistent with the Pa$\alpha$ constraints, CO measurements from \citet{downes98} imply that most of the molecular material for star formation is actually concentrated within the inner $1\farcs3$ (1.1~kpc).  Assuming the IRAM 30~m single-dish CO(1--0) flux of 97~Jy~km~s$^{-1}$ is emitted by a molecular disk with minor and major axes of 300 and 340~pc, respectively, and a height of 23~pc derived from CO rotation curves \citep{downes98}, the CO(1--0) flux implies a nuclear SFR in the range of $\sim$100--350~M$_\odot$~yr$^{-1}$ using the molecular gas density to SFR density relation of \citet{ken98}.  For these calculations, the CO-to-H$_2$ mass conversion ($\alpha_{\rm CO}$) is assumed to be 3--7~M$_\odot$~(K~km~s$^{-1}$~pc$^2$)$^{-1}$ based on the two-phase local velocity gradient models of the circumnuclear molecular gas by \citet{papa12}.  Given that the host-only {\it Chandra} spectrum shows no significant contribution from HMXBs, we will assume all of the X-ray luminosity from HMXBs arises from circumnuclear star formation within 1$\arcsec$.  The CO flux also implies a column density of N$_{\rm H}\sim2.4\times10^{23}$~cm$^{-2}$.  Therefore, the X-ray emission from HMXBs associated with the nuclear starburst would be heavily absorbed and thus not contribute to the X-ray spectrum of Mrk~231 below 2~keV, as discussed by \citet{g02}.


To account for the nuclear starburst, we added a new obscured HMXB component to our modeling.  In addition, deeper {\it Chandra} data show an ionized Fe line at $\sim$6.7~keV, likely arising from the circumnuclear starburst \citep[e.g.,][]{strickland07}.  This line is not self-consistently accounted for by MYTorus, thus we include an additional Gaussian feature to model this line.   To recap, the current model now includes a two-component MEKAL model, an Fe line at 6.7~keV, and a heavily obscured HMXB component for the nuclear star formation (SFR of 140~M$_\odot$~yr$^{-1}$), and a leaky MYTorus component for the AGN emission (see Table~\ref{tab:fits} notes for an equation form of this model).  
For the nuclear starburst component, we assumed values that are consistent with the CO measurements where the absorbing column is assumed to be 2.4$\times10^{23}$~cm$^{-2}$ and the unabsorbed 0.5--8~keV flux is consistent with the \citet{mineo12a} relation for SFR of 140~M$_\odot$~yr$^{-1}$.  All other components in the model are allowed to vary.  

The best-fit values are listed in Column~3 of Table~\ref{tab:fits} and the best-fit spectrum (shown with the model components) is in Figure~\ref{fig:fitsmod}.  The EW of the neutral Fe line compared to the zeroth order power law model is $\sim$0.04~keV and that of the ionized Fe line is $\sim$0.17~keV.  If the 6.7~keV line comes from thermal emission of a starburst, then it should be accompanied by a bremsstrahlung continuum.   However, an additional bremsstrahlung continuum component to the nuclear starburst model contributes very little to the overall fit ($\Delta \chi^2 \sim 0.6$).  This bremsstrahlung component is assumed to have a peak temperature of 4.5~keV in order to produced the line at 6.7~keV with an unabsorbed 0.5--2~keV luminosity equal to the expected value for a 140~M$_\odot$~yr$^{-1}$ SFR and modified by the same column as the HMXB component.  The 6.7~keV line is likely associated with the HMXBs.  The ionized line seems uncommon in high resolution low-mass X-ray binary (LMXB) spectra, but is also rare in HMXBs where only 20\% were detected in the high spectral resolution survey of ten HMXBs conducted by \citet{hmxblines}.  For the HMXB Cen X-3, its 6.7~keV line has EW of 0.02--0.4~keV \citep{cenx3}.  Assuming this is typical of HMXBs, because few high resolution measurements are available, and diluting the EW of the line by 80\% to account for non-detections, the expected EW from HMXBs would be between 0.004--0.1~keV, approximately consistent with the EW range we measured in Mrk~231 (0.06--0.25~keV).  The slightly higher EW measured in Mrk 231 can be partially accounted for by the AGN.

The derived $\Gamma_{\rm AGN}$ of 1.4 is still flatter than the usual range, but not inconsistent with these values.  Fixing the value to 1.8 results in a poorer fit ($\Delta\chi^2 = 42.9$ for a single parameter) and underestimates the flux above 10~keV.  
The intrinsic AGN luminosity remains weak, at only $\sim$0.03\% of the bolometric luminosity.  The line-of-sight column density is Compton-thin with N$_{\rm H}=1.1^{+0.3}_{-0.2}\times10^{23}$~cm$^{-2}$.  
A Compton-thin AGN is consistent with the weak Fe line emission (EW$\sim$0.2~keV).  

The inclination angle of the torus has so far been fixed at 65$^\circ$; this inclination angle implies a sightline through the edge of the obscuring torus.  We also tested geometries where the sightline does not go through the torus at less than 60$^\circ$ (the opening angle assumed by the MYTorus model).  These resulted in poor fits where the model underestimates the emission above 10~keV and overestimates the emission below 5 keV.  If the inclination angle is fixed at 90$^\circ$, the model parameters and fit statistics are approximately the same as those of the best-fit model listed in Column~3 of Table~\ref{tab:fits} except for the column density.  For a geometry looking through the thickest part of the torus, the column density is patchy and Compton-thin (N$_{\rm H} = 1.1^{+0.1}_{-0.1}\times10^{22}$~cm$^{-2}$).   

The inclination angle constraint in MYTorus is mainly driven by the strength of the Fe line and the shape of the hard X-ray spectrum.  Since the Fe line observed in Mrk~231 is so weak and there is no evidence of a Compton-reflection hump, the MYTorus model cannot fully constrain both the inclination angle and the covering factor.  Following \citet{yaqoob12}, we used the MYTorus model to constrain the line-of-sight column density and place an upper limit on the average column density by decoupling the scattered AGN component from that of the direct zeroth order AGN component.  By assuming the best-fit model, we find that the line-of-sight column density is $\sim5.8\times10^{22}$~cm$^{-2}$ and the upper limit to the average column density is $\sim4.0\times10^{24}$~cm$^{-2}$.  
The decoupled column densities are in agreement with the limits we derived from our spectral modeling assuming a toroidal geometry and a sightline cutting through the torus.  

Both \citet{g05} and P13 suggested that the observed flux of Mrk~231 below 2~keV is variable, but \citet{saez12} noted that the intrinsic AGN luminosity of Mrk~231 does not vary.  We re-reduced a 40~ks {\it Chandra} observation from 2003 (ObsID: 04028) presented in \citet{g05} with the same calibration as our 2012 {\it Chandra} data so that the time-dependent ACIS efficiency is taken into account.  Compared to the 2003 {\it Chandra} data, the observed 0.5--2~keV and 2--8~keV fluxes in 2012 have changed by $+3^{+4}_{-45}$\% and $-18^{+3}_{-19}$\%, respectively.  We tested the robustness of our MYTorus results by modeling the {\it NuSTAR} data simultaneously with both the 2003 and 2012 sets of {\it Chandra} data.  Assuming that the intrinsic luminosity of the AGN and the starburst component have not changed as discussed in \S\ref{sec:contam} and \ref{sec:nudata}, we allowed the AGN column density and the AGN Compton-thin luminosity fraction to vary between the 2003 and 2012 datasets in order to explain the observed spectral variability.  The best-fit results are listed in Column 4 of Table~\ref{tab:fits} and shown in Figure~\ref{fig:fits}.  The spectral variability is well-explained by patchy absorption varying between $1.2-2.0 \times 10^{23}$~cm$^{-2}$. 
This model implies an intrinsic 2--10~keV to bolometric luminosity ratio of 0.03\%; thus, Mrk~231 is intrinsically X-ray weak. 

\section{The X-ray Weak AGN in Mrk~231}
\label{sec:discuss}

\subsection{The Flat $\Gamma_{\rm AGN}$ of Mrk~231}
\label{sec:gamma}

The physically motivated MYTorus model appears to provide an excellent description of our broadband X-ray spectrum of Mrk 231.  The best-fit model implies a Compton-thin absorbing column (N$_{\rm H}\sim1.2\times10^{23}$~cm$^{-2}$).  
  Although $\Gamma_{\rm AGN}$ of 1.4 is not very much lower than the typical range seen in AGN and the upper limit is within the expected range, the derived value is an artificial constraint as it is limited by the lower bound (1.4) of the MYTorus table model.  It is possible that the true value of $\Gamma_{\rm AGN}$ is even lower than 1.4.  As discussed in \S\ref{sec:multi}, the flatness of the Mrk~231 spectrum below 10~keV has been assumed to be due to a strong reflection component, and therefore the absorbing column is Compton-thick.  It is not until our {\it NuSTAR} observations that the lack of a distinct reflection hump above 10~keV is definitively shown.  

One explanation for the flat $\Gamma_{\rm AGN}$ is that Mrk~231 is so obscured that even MYTorus, with an N$_{\rm H}$ upper bound of 10$^{25}$~cm$^{-2}$, is ineffective in fully describing the broadband spectrum because the actual column is above the upper bound.  This is unlikely.  If the intrinsic AGN luminosity were higher, then in order to produce the observed spectrum, a thicker absorbing column is required.  This in turn implies that more of the direct AGN emission is Compton scattered by the torus.  Since the observed soft X-ray spectrum is almost completely accounted for by the starburst component, there is little leeway in the observed spectrum to accomodate more scattered emission below 2~keV.  If the absorbing column is above 10$^{25}$~cm$^{-2}$ and has a high covering factor ($\approx4\pi$), then it is unlikely that a bright optical nucleus would be detected.  
It is also possible that the assumed half-opening angle of MYTorus (60$^\circ$) is incorrect.  We examined this possibility by applying a version of the MYTorus model\footnote{These models were provided by T. Yaqoob with new calculations only relevant to the continuum (i.e. the transmitted and Compton-scattered components), so the strength of the Fe line is not taken into account.  This version of the model may not be reliable below $\sim$3~keV. } available to us that assumes a half-opening angle of 37$^\circ$ to our data.  The model parameters from this test are consistent with those derived from the standard MYTorus model and a sightline going through the torus.  

Another possibility is that the flat $\Gamma_{\rm AGN}$ is an intrinsic property of the AGN in Mrk~231.  In ULIRG-type sources, black hole growth occurs most rapidly after coalescence and during the obscured phase \citep[e.g.,][]{martinez05, teng10}.  The high accretion rate of the black hole also drives feedback which disperses the obscuring material. In most cases, a steep $\Gamma_{\rm AGN}$ is associated with high Eddington ratios \citep[e.g.,][]{shemmer08, brightman13}.   In this case, perhaps a disturbed corona, a product of the ongoing high accretion rate of the AGN and powerful outflow, produces a flatter than typical photon index.  Galactic X-ray binaries in the ``ultra soft state'' are characterized by a high Eddington ratio but low X-ray luminosity \citep[e.g.,][]{done07}.
Mrk~231 being an AGN in a rare ultra soft state is consistent with the lack of intrinsic variability and the low X-ray-to-optical power law slope ($\alpha_{\rm OX}$, see below).  However, $\Gamma_{\rm AGN}$ for the ultra soft state would be $\sim$2 \citep[e.g.,][]{done07}, steeper than our measurement.  A flat $\Gamma$ ($< 1.4$) implies the lack of low energy seed photons that can be Compton up-scattered to higher energies \citep[e.g.,][]{zdziarski88, fabian88}; this does not seem likely for Mrk~231 given its UV and optical properties.  Furthermore, given the super-Eddington accretion rate measured for Mrk~231 (see \S~\ref{sec:outlier}), a standard scenario of an advection-dominated accretion flow \citep[e.g.,][]{narayan94, narayan95}, where the flat $\Gamma_{\rm AGN}$ mimics a thermal bremsstrahlung spectrum \citep[e.g.,][]{bremspl}, is unlikely.

It is also possible that Mrk~231 is in fact a radio-loud object.  The radio-loud BAL quasar PG~1004+130 is thought to be a jetted source that is likely to be intrinsically X-ray weak \citep{luo13}.  X-ray spectra of radio-loud sources can have relatively flat slopes \citep[e.g., NGC~4261;][]{sambruna}.  If the AGN corona is radiatively inefficient and is feeding a jet, then the X-ray emission may be dominated by inverse Compton, resulting in a flat $\Gamma_{\rm AGN}$ \citep[e.g.,][]{markoff05}.  Recent Very Long Baseline Array monitoring of Mrk~231 by \citet{reynolds13} discovered a blazar-like radio flare in Mrk~231.  The first epoch of our {\it NuSTAR} data was taken before the most recent radio flare while the second epoch was taken soon after the peak of the flare.  If the flat $\Gamma_{\rm AGN}$ is related to the launching of a jet or a flare, then one would likely expect changes in the X-ray slope as flares come and go.  We find no significant changes in $\Gamma_{\rm AGN}$ between the two epochs of {\it NuSTAR} data nor significant variability in $\Gamma_{\rm AGN}$ and X-ray flux in historic X-ray data of Mrk~231.  In addition, the expected X-ray flux from synchrotron self-Compton derived from the radio flux density of the flare is only $\sim$10$^{-21}$~erg~s$^{-1}$~cm$^{-2}$, far below the measured X-ray flux.  Therefore, it is unlikely that the flares contribute significantly to the observed X-ray properties of Mrk~231.

\subsection{Mrk~231 is an Outlier Among Quasars}
\label{sec:outlier}

Regardless of the reason behind the flat photon index of Mrk~231, the absorption-corrected luminosities from all of our modeling imply that the total intrinsic AGN luminosity for Mrk~231 in 0.5--30~keV is $\sim1.0\times10^{43}$~erg~s$^{-1}$.  Compared to L$_{\rm bol, AGN}$ of $1.1\times10^{46}$~erg~s$^{-1}$, the 2--10~keV X-ray luminosity is 0.03--0.05\% of the AGN bolometric luminosity.  Even if we consider that the AGN only contributes to $\sim$30\% of the bolometric luminosity as derived by the modeling of the far-infrared SED \citep[L$_{\rm bol, AGN} = 4.7\times10^{45}$~erg~s$^{-1}$;][]{farrah03}, the ratio between 2--10~keV luminosity and L$_{\rm bol, AGN}$ is still $\sim$0.08\%.  The derived AGN fraction is much lower than the value expected for this type of source.  For comparison, Seyfert~1 galaxies and radio-quiet quasars sampled by \citet{elvis} for which X-ray data were available have X-ray to bolometric ratios of $\sim$2 to 15\%, with the most luminous objects typically having the lowest ratios.  However, this does not seem to be unusual for ULIRG-type sources.  \citet{teng10} found that Seyfert~1-like ULIRGs have mean 2--10~keV to bolometric ratios of $\sim$0.1\% compared to the mean of $\sim$3\% for PG~quasars.   

Recent studies by \citet{vf07, vf09} and \citet{lusso10, lusso12} have independently confirmed that AGN with the highest Eddington ratios require the highest bolometric corrections.  Assuming the dynamically derived black hole mass of $1.7^{+4.0}_{-1.2}\times10^7$~M$_\odot$ using stellar velocity dispersions measured from near-infrared spectra \citep{dasyra}\footnote{There is some uncertainty to the derived black hole mass and therefore the Eddington ratio.  Using H$\beta$ line widths, \citet{kawakatu07} found the black hole in Mrk~231 is $8.7\times10^7$~M$_\odot$ with an uncertainty that is within a factor of $\sim$3.  Therefore, the H$\beta$ value is broadly consistent with the \citet{dasyra} value that we have assumed.  The H$\beta$ spectra were taken using a 3$\arcsec$ aperture, so there may be contamination from the host.  Thus, we consider the H$\beta$-derived black hole mass an upper limit, leading to an Eddington ratio lower limit of $\sim$1.1.} and L$_{\rm bol, AGN}$ of $1.1\times10^{46}$~erg~s$^{-1}$, the Eddington ratio of the AGN in Mrk~231 is 5.0$^{+11.3}_{-3.5}$, assuming isotropic emission.  The 2--10~keV to bolometric luminosity ratio for an AGN with such an Eddington ratio is expected to be $\sim$0.3\% \citep{lusso10, lusso12} to $\sim$0.7\% \citep{vf09}.  
However, the correlation has a large dispersion and lacks sampling at the very X-ray weak end. Therefore, it is difficult to assess whether the significant X-ray weakness can be explained simply by super-Eddington accretion.  

\begin{figure}[ht]
\centering
\includegraphics[width=2.5in, angle=270]{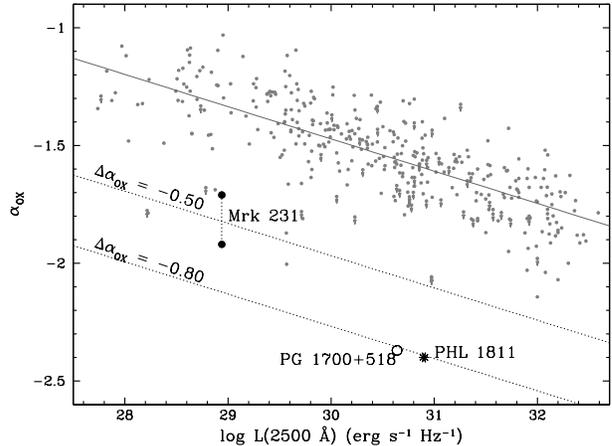}
\caption{$\alpha_{\rm OX}$ versus the 2500~\AA~monochromatic luminosity for Mrk~231.  The 2500~\AA~flux of Mrk~231 has been corrected for dust reddening based on modeling of the UV continuum by \citet{vei13a}.  The range of $\alpha_{\rm OX}$ for Mrk~231 plotted here represent the observed total 2~keV flux (--1.92) and the absorption-corrected direct AGN 2~keV flux (--1.71).  Also included are the X-ray weak AGN PHL~1811 \citep{phl1811a, phl1811b} and BAL quasar PG~1700+518 \citep{luo13}.  The grey points are normal quasars from \citet{steffen06} and the grey line is their $\alpha_{\rm OX}$-optical luminosity relation for radio-quiet quasars.  The dotted lines mark deviations of --0.50 and --0.80 from the \citet{steffen06} relation.  The X-ray weakness of Mrk~231 is distinct compared to the ``normal'' AGN studied by \citet{steffen06}.  }  
\label{fig:alphaox}
\end{figure}

Although rare, Mrk~231 is not the only intrinsically X-ray weak AGN known.  The X-ray-to-optical power law slope ($\alpha_{\rm OX}$; Figure~\ref{fig:alphaox}) of Mrk~231 is not as weak as that of the known X-ray weak AGN PHL~1811, an X-ray variable AGN with a steep photon index \citep[$\Gamma_{\rm AGN}=2.3$,][]{phl1811a, phl1811b}.  X-ray stacking analysis of a sample of radio-quiet high-redshift ($z \sim 2.2$) analogs of PHL~1811 shows that, on average, they have $\Gamma_{\rm eff} \sim 1.1$ \citep{wu11}.  These sources appear to have large blueshifted C~IV absorption line equivalent widths, likely due to winds dominating their broad emission line regions \citep{wu11}.  Since PHL~1811 itself has a soft spectrum of $\Gamma_{\rm AGN} = 2.3$, its analogs are plausibly expected to have similarly soft X-ray spectra.  Therefore, the flat effective X-ray photon indices measured from the analogs are interpreted as indications of heavy obscuration and probably even reflection-dominated spectra.  To place Mrk~231 in the context of other LoBAL quasars and X-ray weak AGN, we redshifted our best-fit model of Mrk~231 to the redshift of the other radio-quiet LoBAL quasar, PG~1700+518, observed by {\it NuSTAR} \citep[$z=0.292$;][]{luo13}.  The effective photon index of the redshifted spectrum ($\Gamma_{\rm eff}\sim0.9$) is similar to that measured in PG~1700+518 ($\Gamma_{\rm eff}\sim0.5$).  
Mrk~231 seems to be a local example of an X-ray faint LoBAL \citep[17--40\% of which are intrinsically X-ray weak;][]{luo13} that innately possesses a weak X-ray continuum.  

The X-ray weakness of the ionizing continuum is also manifested in the lack of [O IV] (25.89~$\mu$m), [Ne V] (14.32, 24.32~$\mu$m), and [Ne VI] (7.65~$\mu$m) line emission from Mrk~231 \citep{armus, farrah07, vei09a}.  These lines arise in the narrow line region, located above the torus, and are thus less likely to be affected by obscuration.  The [O IV], [Ne V] and [Ne VI] lines require sufficient high energy photons ($>$77~eV) to be produced.  The weak continuum of Mrk~231 becomes more evident if we compare it to another local ($z = 0.043$) AGN ULIRG, IRAS~F05189--2524.  Both ULIRGs are in the same merger stage \citep{vei09b}, have approximately the same dynamically derived black hole masses \citep{dasyra}, have nearly identical infrared SEDs \citep{armus}, and have almost the same L$_{\rm bol, AGN}$ \citep{vei09a}.  However, they have very different derived Eddington ratios ($\sim$1 versus $\sim$5).  Given the uncertainties in the measured black hole masses and Eddington ratios, this is only a qualitative comparison between Mrk~231 and IRAS~F05189--2524.  The comparison is valid because the black hole masses were measured using the same method and thus have similar systemic measurement errors.   IRAS~F05189--2524 has strong [O IV], [Ne V], and [Ne VI] emission lines and a strong intrinsic X-ray continuum \citep{ptak03, teng09} while Mrk~231 is the opposite.  Although we only have a limited sample, the distinction between these two sources may explain the large scatter of [O IV]-to-bolometric luminosity relations observed in nearby AGN \citep[e.g.,][]{melendez08, rigby09, weaver10}.  The intrinsic X-ray weakness of Mrk~231 is likely associated with the super-Eddington accretion rate of the black hole that also drives its powerful outflows \citep{rv13}.  

\section{Conclusions}
\label{sec:summary}

We have obtained the most sensitive broadband (0.5--30~keV) spectrum of Mrk~231, the closest BAL quasar and an ULIRG,  to date.  {\it NuSTAR} detects Mrk~231 at a fainter level above 10~keV than previous observations with poorer angular resolution by {\it BeppoSAX} PDS and {\it Suzaku} PIN.  Two epochs of {\it NuSTAR} data separated by seven months show no evidence of hard X-ray variability so contamination is the likely cause of the discrepancy between {\it NuSTAR} and previous, non-focusing hard X-ray observations.  

Analysis of contemporaneous {\it Chandra} and {\it NuSTAR} broadband spectra find significant X-ray emission from a powerful circumnuclear starburst.  The direct AGN emission is absorbed and scattered by a patchy torus that is Compton-thin with N$_{\rm H}\sim1.2^{+0.3}_{-0.3}\times10^{23}$~cm$^{-2}$.  The Compton-thin absorption is compatible with the apparent weak Fe~K emission (EW$\sim$0.2~keV).

Mrk~231 appears intrinsically X-ray weak.  The absorption-corrected 2--10~keV luminosity is $\lesssim0.1$\% of the AGN bolometric luminosity with $\alpha_{\rm OX}\sim-1.7$.  A reason for the low X-ray luminosity could be that the AGN is in a rare ultra soft state similar to those seen in X-ray binaries, though this is inconsistent with the measured $\Gamma_{\rm AGN}$.  It is unclear why the AGN is intrinsically X-ray weak, but there are not very many known examples of such sources as points of comparison.  The weak ionizing continuum explains the lack of mid-infrared [O IV], [Ne V], and [Ne VI] fine structure emission lines which are present in sources with otherwise similar AGN properties.  
The intrinsic X-ray weakness may be a result of the super-Eddington accretion occurring in the nucleus of the ULIRG.  These results on Mrk~231 are consistent with its being a merger remnant emerging from its dusty cocoon through a powerful wind event.

\acknowledgements We are grateful to the anonymous referee for providing useful comments which improved our manuscript.  We thank Wayne Baumgartner, Bret Lehmer, Richard Mushotzky, Jeremy Schnittman, Tahir Yaqoob, and Andreas Zezas for useful discussions.  We would also like to thank Lee Armus who provided useful comments in the early planning phase of the {\it NuSTAR} ULIRG program.  We also thank Roberto Maiolino, David Rupke, and Eckhard Sturm who are co-investigators of the {\it Chandra} program.  This work was supported under NASA Contract No. NNG08FD60C, and  made use of data from the {\it NuSTAR} mission, a project led by  the California Institute of Technology, managed by the Jet Propulsion  Laboratory, and funded by the National Aeronautics and Space  Administration. We thank the {\it NuSTAR} Operations, Software and  Calibration teams for support with the execution and analysis of  these observations.  This research has made use of the {\it NuSTAR}  Data Analysis Software (NuSTARDAS) jointly developed by the ASI  Science Data Center (ASDC, Italy) and the California Institute of  Technology (USA).  The scientific results reported in this article are based in part on observations made by the {\it Chandra X-ray Observatory} and data obtained from the {\it Chandra} Data Archive published previously in cited articles.  This work, in part, made use of observations obtained with {\it XMM-Newton}, an ESA science mission with instruments and contributions directly funded by ESA Member States and the USA (NASA).  We made use of the NASA/IPAC Extragalactic Database (NED), which is operated by the Jet Propulsion Laboratory, Caltech, under contract with NASA.  S.H.T. is supported by a NASA Postdoctoral Program (NPP) Fellowship.  W.N.B. and B.L. acknowledge support by California Institute of Technology (Caltech) NuSTAR subcontract 44A-1092750 and NASA ADP grant NNX10AC99G. F.E.B acknowledges support from Basal-CATA (PFB-06/2007) and CONICYT-Chile (under grants FONDECYT 1101024 and Anillo ACT1101).  A.C. acknowledges support from ASI/INAF grant I/037/12/0-011/13.   P.G. acknowledges support from STFC grant reference ST/J003697/1.

{\it Facilities:} \facility{{\it NuSTAR}, {\it Chandra}}.

\begin{turnpage}
\begin{deluxetable}{lcccl}
\tablecolumns{5}
\tabletypesize{\tiny}
\setlength{\tabcolsep}{0.01in}
\tablecaption{Best-fit Parameters}
\tablewidth{0pt}
\tablehead{\colhead{Model} & \colhead{Partial Covering} & \colhead{MYTorus\tablenotemark{a}} & \colhead{MYTorus\tablenotemark{a}} &\colhead{Comment on}\\
\colhead{Parameter} & \colhead{Absorber}&\colhead{(2012)}& \colhead{(2003+2012)} &\colhead{Parameter}\\
\colhead{(1)}&\colhead{(2)}&\colhead{(3)}&\colhead{(4)}&\colhead{(5)}
}
\startdata
B/A&0.97$^{+0.07}_{-0.07}$&0.97$^{+0.07}_{-0.07}$&0.97$^{+0.07}_{-0.07}$&FPMB-A cross-normalization\\
CXO/A&1.00$^{+0.10}_{-0.10}$&0.97$^{+0.11}_{-0.10}$&0.90$^{+0.09}_{-0.08}$& {\it Chandra}-FPMA cross-normalization\\
$k$T [keV]&0.82$^{+0.07}_{-0.09}$&0.25$^{+0.05}_{-0.06}$, 0.93$^{+0.11}_{-0.10}$ &0.26$^{+0.06}_{-0.05}$, 0.87$^{+0.09}_{-0.08}$&MEKAL gas temperature from the starburst\\
Abs. 1 [$\times 10^{22}$~cm$^{-2}$]\tablenotemark{b}&6.57$^{+1.82}_{-1.40}$&11.2$^{+2.9}_{-2.4}$&19.4$^{+5.7}_{-4.4}$\tablenotemark{d}, 9.5$^{+2.3}_{-1.9}$\tablenotemark{e} & neutral absorber 1\\
cf 1&0.82$^{+0.12}_{-0.03}$& 1 (f) &1 (f) & covering factor 1\\
Abs. 2 [$\times 10^{22}$~cm$^{-2}$]\tablenotemark{c}&\nodata& 24 (f)& 24 (f) &neutral absorber 2\\
$\Gamma_{\rm HMXB}$&1.1 (f)&1.1 (f)& 1.1 (f) &HMXB cutoff power law index with cutoff energy at 10~keV\\
$\Gamma_{\rm AGN}$&1.39$^{+0.10}_{-0.11}$&1.40$^{+0.04}_{...}$&1.40$^{+0.03}_{...}$&AGN power law index (MYTorus lower limit fixed at 1.4)\\
Inc [$^\circ$]&\nodata&65 (f) &65 (f)&inclination angle\\
E$_{\rm line}$[keV]&6.63$^{+0.07}_{-0.06}$&6.67$^{+0.03}_{-0.10}$&6.66$^{+0.04}_{-0.04}$&narrow ionized Fe line energy (Fe K$\alpha$ self consistent within MYTorus)\\
EW$_{\rm line}$[keV]&0.168$^{+0.093}_{-0.095}$&0.146$^{+0.121}_{-0.075}$&0.166$^{+0.119}_{-0.070}$&ionized Fe line equivalent width\\
Const. (C-thin)&\nodata&0.19$^{+0.04}_{-0.03}$&0.23$^{+0.04}_{-0.03}$\tablenotemark{d}, 0.20$^{+0.04}_{-0.04}$\tablenotemark{e}&Compton-thin fraction\\
$f_{0.5-2}$ [$\times 10^{-13}$ erg~s$^{-1}$~cm$^{-2}$]&$1.07^{+1.76}_{-0.03}$&1.16$^{+0.05}_{-0.10}$&$1.19^{+0.04}_{-0.10}$\tablenotemark{d}, $1.16^{+0.04}_{-0.06}$\tablenotemark{e}&observed 0.5--2 keV flux\\
$f_{2-10}$ [$\times 10^{-13}$ erg~s$^{-1}$~cm$^{-2}$]&$9.42^{+3.59}_{-0.89}$&9.20$^{+0.31}_{-2.30}$&$7.69^{+0.20}_{-1.75}$\tablenotemark{d}, $8.72^{+0.27}_{-1.94}$\tablenotemark{e}&observed 2--10 keV flux\\
$f_{10-30}$ [$\times 10^{-12}$ erg~s$^{-1}$~cm$^{-2}$]&$1.82^{+0.22}_{-0.37}$&1.77$^{+0.01}_{-0.62}$&$1.75^{+0.01}_{-0.56}$&observed 10--30 keV flux\\
L$_{\rm MEKAL}$ [erg~s$^{-1}$]&$1.51 \times 10^{41}$&$2.30\times10^{41}$&$2.35 \times 10^{41}$&intrinsic MEKAL 0.5--30~keV luminosity\\
L$_{\rm HMXB}$ [erg~s$^{-1}$]&$8.53 \times 10^{39}$&$8.57\times10^{41}$&$8.57\times 10^{41}$&intrinsic HMXB 0.5--30~keV luminosity\\
L$_{0}$ [erg~s$^{-1}$]&$1.40\times 10^{43}$&$1.13\times10^{43}$&$1.01 \times 10^{43}$&intrinsic AGN 0.5--30 keV luminosity\\
L$_{0}$ (2--10~keV) [erg~s$^{-1}$]&$4.90\times 10^{42}$&$3.94\times10^{42}$&$3.84 \times 10^{42}$&intrinsic AGN 2--10 keV luminosity\\
L$_{\rm C-thin}$ [erg~s$^{-1}$]&\nodata&$2.18\times10^{42}$&$2.38 \times 10^{42}$\tablenotemark{d}, $2.23 \times 10^{42}$\tablenotemark{e}&Compton-thin AGN 0.5--30~keV luminosity\\
L$_{\rm AGN}$/L$_{\rm bol, AGN}$ [\%]&0.05&0.03&0.03 &2--10~keV X-ray-to-bolometric luminosity ratio for the AGN\\
\hline\\
$\chi^2$/d.o.f.&258.3/258&249.7/258&390.2/380&goodness of fit\\
\enddata
\tablecomments{
Col.(1): Model parameter for each fit.  Col.(2):  A single partial covering absorber model.  Col.(3): MYTorus fits to the contemporaneous {\it Chandra} and {\it NuSTAR} data.  Col.(4): Same as column (3), but fitting 2003 and 2012 {\it Chandra} data simultaneously.  Col.(5):  Comments on model parameter.
}
\tablenotetext{a}{Best-fit model: Const.$\times$N$_{\rm H, Galactic}$(MEKAL$_{1}$+MEKAL$_{2}$+Abs$_{\rm nuclear~HMXB}\times$cutoffPL$_{\rm nuclear~HMXB}$+Line(6.7~keV)+MYTorus$\times$PL$_{\rm AGN}$+Const.$_{\rm C-thin}\times$PL$_{\rm AGN}$)}
\tablenotetext{b}{Absorbing column of the torus in the MYTorus fits.}
\tablenotetext{c}{Absorbing column of the HMXB component in the MYTorus fits.}
\tablenotetext{d}{2003 values.}
\tablenotetext{e}{2012 values.}
\label{tab:fits}
\end{deluxetable}
\end{turnpage}


\begin{thebibliography}{}
\bibitem[Adams \& Weedman(1972)]{aw72} Adams, T.F. \& Weedman, D.W. 1972, ApJ, 173, L109
\bibitem[Armus et al.(2007)]{armus} Armus, L., et al. 2007, ApJ, 656, 148
\bibitem[Baumgartner et al.(2013)]{bat} Baumgartner, W.H., et al. 2013, ApJS, 207, 19
\bibitem[Becker et al.(1995)]{first} Becker, R.H., White, R.L., \& Helfand, D.J., 1995, ApJ, 450, 550
\bibitem[Boksenberg et al.(1977)]{boksenberg} Boksenberg, A. et al. 1977, MNRAS, 178, 451
\bibitem[Braito et al.(2004)]{braito} Braito, V., et al. 2004, A\&A, 420,79 (B04)
\bibitem[Brightman et al.(2013)]{brightman13} Brightman, M. et al. 2013, MNRAS, 433, 2485
\bibitem[Carilli et al.(1998)]{carilli98} Carilli, C. L., Wrobel, J. M., \& Ulvestad, J. S. 1998, AJ, 115, 928
\bibitem[Dai et al.(2012)]{dai12} Dai, X., Shankar, F., \& Sivakoff, G.R., 2012, ApJ, 757, 180
\bibitem[Dasyra et al.(2006)]{dasyra} Dasyra, K., et al. 2006, ApJ, 651, 835
\bibitem[Di Matteo et al.(2005)]{dimatteo} Di Matteo et al. 2005, Nature, 433, 604
\bibitem[DiPompeo et al.(2013)]{dipompeo13} DiPompeo, M.A., Brotherton, M.S., \& De Breuck, C. 2013, MNRAS, 428, 1565
\bibitem[Dickey \& Lockman(1990)]{nh} Dickey, J.M. \& Lockman, F.J. 1990, ARA\&A, 28, 215
\bibitem[Done et al.(2007)]{done07} Done, C., Gierli\'{n}ski, M., \& Kubota, A. 2007, A\&ARv, 15, 1
\bibitem[Downes \& Solomon(1998)]{downes98} Downes, D. \& Solomon, P.M. 1998, ApJ, 507, 615
\bibitem[Elvis et al.(1994)]{elvis} Elvis, M., et al. 1994, ApJS, 95, 1
\bibitem[Fabian et al.(1988)]{fabian88} Fabian, A.C., Done, C., \& Ghisellini, G., 1988, MNRAS, 232, 21
\bibitem[Farrah et al.(2003)]{farrah03} Farrah, D., Afonso, J., Efstathiou, A., Rowan-Robinson, M., Fox, M., \& Clements, D. 2003, MNRAS, 343, 585
\bibitem[Farrah et al.(2007)]{farrah07} Farrah, D., et al. 2007, ApJ, 667, 149
\bibitem[Feruglio et al.(2010)]{feruglio} Feruglio, C., Maiolino, R., Piconcelli, E., Menci, N., Aussel, H., Lamastra, A.,
\& Fiore, F. 2010, A\&A, 518, L155
\bibitem[Fischer et al.(2010)]{fischer} Fischer, J., Sturm, E., Gonzalez-Alfonso, E., et al. 2010, A\&A, 518, L41
\bibitem[Gallagher et al.(2002)]{g02} Gallagher et al. 2002, ApJ, 569, 655
\bibitem[Gallagher et al.(2005)]{g05} Gallagher et al. 2005, ApJ, 633, 71
\bibitem[Gallagher et al.(2006)]{g06} Gallagher et al. 2006, ApJ, 644, 709
\bibitem[Gebhardt et al.(2000)]{gebhardt00} Gebhardt, K. et al. 2000, ApJ, 539, L13
\bibitem[Gibson et al.(2009)]{gibson09} Gibson, R.R. et al. 2009, ApJ, 692, 758
\bibitem[Giustini et al.(2008)]{giustini08} Giustini, M., Cappi, M., \& vignali, C. 2008, A\&A, 491, 425
\bibitem[Gonz\'{a}lez-Alfonso et al.(2014)]{ga14} Gonz\'{a}lez-Alfonso, E. et al. 2014, A\&A, 561, 27
\bibitem[Grupe et al.(2003)]{grupe03} Grupe, D., Mathur, S., \& Elvis, M. 2003, AJ, 126, 1159
\bibitem[Harrison et al.(2013)]{nustar} Harrison, F.A. et al. 2013, ApJ, 770, 103
\bibitem[Hinshaw et al.(2009)]{cosmo} Hinshaw, G., Weiland, J.L., Hill, R.S., et al. 2009, ApJS, 180, 225
\bibitem[Hopkins et al.(2008)]{hopkins08} Hopkins, P.F. et al. 2008, ApJ, 175, 356
\bibitem[Kawakatu et al.(2007)]{kawakatu07} Kawakatu, N., Imanishi, M., \& Nagao, T. 2007, ApJ, 661, 660
\bibitem[Kennicutt(1998)]{ken98} Kennicutt, R.C. 1998, ApJ, 498, 541
\bibitem[Kim \& Sanders(1998)]{ks98} Kim, D.-C. \& Sanders, D.B. 1998, ApJS, 119, 41
\bibitem[Kim et al.(2007)]{lognlogs} Kim, M., Wilkes, B.J., Kim, D.-W., Green, P.J., Barkhouse, W.A., Lee, M.G., Silverman, J.D., \& Tananbaum, H.D. 2007, ApJ, 659, 29
\bibitem[Koss et al.(2013)]{koss} Koss, M. et al. 2013, ApJ, 765, L26
\bibitem[Lehmer et al.(2010)]{lehmer} Lehmer, B.D. et al. 2010, ApJ, 724, 559
\bibitem[Leighly et al.(2007a)]{phl1811a} Leighly, K.M., Halpern, J.P., Jenkins, E.B. et al. 2007b, ApJ, 663, 103
\bibitem[Leighly et al.(2007b)]{phl1811b} Leighly, K.M., Halpern, J.P., Jenkins, E.B., \& Casebeer, D. 2007, ApJS, 173, 1
\bibitem[Luo et al.(2013)]{luo13} Luo, B. et al. 2013, ApJ, in press (arXiv:1306.3500)
\bibitem[Lusso et al.(2010)]{lusso10} Lusso, E. et al. 2010, A\&A, 512, 34
\bibitem[Lusso et al.(2012)]{lusso12} Lusso, E. et al. 2012, MNRAS, 425, 623
\bibitem[Lutovinov et al.(2005)]{hmxb2} Lutovinov, A., Revnivtsev, M., Gilfanov, M., Shtykovskiy, P., Molkov, S., \& Sunyaev, R. 2005, A\&A, 444, 821
\bibitem[Lynds(1967)]{lynds67} Lynds, C.R. 1967, ApJ, 147, 396
\bibitem[Maloney \& Reynolds(2000)]{malreynolds} Maloney, P.R. \& Reynolds, C.S. 2000, ApJ, 545, L23
\bibitem[Markoff et al.(2005)]{markoff05} Markoff, S., Nowak, M.A., \& Wilms, J. 2005, ApJ, 635, 1203
\bibitem[Mart\'{i}nez-Sansigre et al.(2005)]{martinez05} Mart\'{i}nez-Sansigre, A., Rawlings, S., Lacy, M., Fadda, D., Marleau, F.R., Simpson, C., Willott, C.J., \& Jarvis, M.J., 2005, Nature, 436, 666
\bibitem[Melendez et al.(2008)]{melendez08} Melendez, M., Kraemer, S.B., Schmitt, H.R., et al. 2008, ApJ, 689, 95
\bibitem[Mineo et al.(2012a)]{mineo12a} Mineo, S., Gilfanov, M., \& Sunyaev, R. 2012, MNRAS, 419, 2095
\bibitem[Mineo et al.(2012b)]{mineo12b} Mineo, S., Gilfanov, M., \& Sunyaev, R. 2012, MNRAS, 426, 1870
\bibitem[Morganti(2011)]{morganti11} Morganti, R. 2011, arXiv:1112.5093
\bibitem[Murphy \& Yaqoob(2009)]{mytorus} Murphy, K.D. \& Yaqoob, T. 2009, MNRAS, 397, 1549
\bibitem[Murray et al.(1995)]{murray95} Murray, N., Chiang, J., Grossman, S.A., \& Voit, G.M. 1995, ApJ, 451, 498
\bibitem[Nagase(1989)]{hmxb} Nagase, F. 1989, PASJ, 41, 1
\bibitem[Naik et al.(2011)]{cenx3} Naik, S., Paul, B., \& Ali, Z. 2011, ApJ, 737, 79
\bibitem[Nandra \& Pounds(1994)]{nandra} Nandra, K. \& Pounds, K.A., 1994, MNRAS, 268, 405
\bibitem[Nandra et al.(2007)]{nandra07} Nandra, K. et al., 2007, MNRAS, 382, 194
\bibitem[Narayan \& Yi(1994)]{narayan94} Narayan, R. \& Yi, I., 1994, ApJ, 428, L13 
\bibitem[Narayan \& Yi(1995)]{narayan95} Narayan, R. \& Yi, I., 1995, ApJ, 452, 710
\bibitem[Ogle et al.(1999)]{ogle99} Ogle, P.M., Cohen, M.H., Miller, J.S., et al. 1999, ApJS, 125, 1
\bibitem[Osterbrock(1978)]{osterbrock78} Osterbrock, D.E. 1978, Phys. Scripta, 17, 137
\bibitem[Persic \& Raphaeli(2002)]{persic02} Persic, M. \& Raphaeli, Y. 2002, A\&A, 382, 843
\bibitem[Piconcelli et al.(2013)]{piconcelli} Piconcelli et al. 2013, MNRAS, 428, 1185 (P13)
\bibitem[Papadopoulos et al.(2012)]{papa12} Papadopoulos, P.P., van der Werf, P.P., Xilouris, E.M., Isaak, K.G., Gao, Y., \& M\"{u}hle, S., 2012, MNRAS, 426, 2601
\bibitem[Pozdnyakov et al.(1983)]{bremspl} Pozdnyakov, L.A., Sobol, I.M., \& Syunyaev, R.A., 1983, Astrophysics and Space Physics Reviews, 2, 189
\bibitem[Proga et al.(2000)]{proga00} Proga, D., Stone, J.M., \& Kallman, T.R. 2000, ApJ, 543, 686
\bibitem[Ptak et al.(2003)]{ptak03} Ptak, A. et al. 2003, ApJ, 592, 782
\bibitem[Reeves \& Turner(2000)]{reeves} Reeves, J.N. \& Turner, M.I.J., 2000, MNRAS, 316, 234
\bibitem[Reynolds et al.(2013)]{reynolds13} Reynolds, C., Punsly, B., O'Dea, C.P., \& Hurley-Walker, N., 2013, ApJ, 776, L21
\bibitem[Rigby et al.(2009)]{rigby09} Rigby, J.R., Diamond-Stanic, A.M., \& Aniano, G., 2009, ApJ, 700, 1878
\bibitem[Rupke et al.(2005)]{rupke05} Rupke, D.S., Veilleux, S., \& Sanders, D.B. 2005, ApJ, 632, 751
\bibitem[Rupke \& Veilleux(2011)]{rv11} Rupke, D.N.S. \& Veilleux, S., 2011, ApJ, 729, L27
\bibitem[Rupke \& Veilleux(2013)]{rv13} Rupke, D.N.S. \& Veilleux, S., 2013, ApJ, 768, 75
\bibitem[Saez et al.(2012)]{saez12} Saez, C., Brandt, W.N., Gallagher, S.C., Bauer, F.E. \& Garmire, G.P., 2012, ApJ, 759, 42
\bibitem[Sambruna et al.(2003)]{sambruna} Sambruna, R.M., Gliozzi, M., Eracleous, M., Brandt, W.N., \& Mushotzky, R., 2003, ApJ, 586, L37
\bibitem[Sanders et al.(1988)]{sanders} Sanders, D.B. et al. 1988, ApJ, 328, L35
\bibitem[Sanders et al.(2003)]{rbgs} Sanders, D.B., Mazzarella, J. M., Kim, D.-C., Surace, J.A., \& Soifer, B.T., 2003, AJ, 126, 1607
\bibitem[Shemmer et al.(2005)]{shemmer05} Shemmer, O., Brandt, W.N., Gallagher, S.C., et al. 2005, AJ, 130, 2522
\bibitem[Shemmer et al.(2008)]{shemmer08} Shemmer, O., Brandt, W.N., Netzer, H., Maiolino, R., \& Kaspi, S., 2008, ApJ, 682, 81
\bibitem[Steffen et al.(2006)]{steffen06} Steffen, A.T., Strateva, I., Brandt, W.N., et al. 2006, AJ, 131, 1163
\bibitem[Strickland \& Heckman(2007)]{strickland07} Strickland, D.K. \& Heckmand, T.M. 2007, ApJ, 658, 258
\bibitem[Surace et al.(1998)]{surace98} Surace, J.A., Sanders, D.B., Wacca, W.D., Veilleux, S., \& Mazzarella, J.M. 1998, ApJ, 492, 116
\bibitem[Teng et al.(2009)]{teng09} Teng, S.H. et al. 2009, ApJ, 691, 261
\bibitem[Teng \& Veilleux(2010)]{teng10} Teng, S.H. \& Veilleux, S. 2010, ApJ, 725, 1848
\bibitem[Teng et al.(2013)]{teng13} Teng, S.H., Veilleux, S., \& Baker, A.J. 2013, ApJ, 765, 95
\bibitem[Torrejon et al.(2010)]{hmxblines}  Torrejon, J.M., Schulz, N.S., Nowak, M.A., \& Kallman, T.R. 2010, ApJ, 715, 947
\bibitem[Ulvestad et al.(1999)]{ulvestad} Ulvestad, J.S., Wrobel, J.M., \& Carilli, C.L., 1999, 516, 127
\bibitem[Vasudevan \& Fabian(2007)]{vf07} Vasudevan, R.V. \& Fabian, A.C. 2007, MNRAS, 381, 1235
\bibitem[Vasudevan \& Fabian(2009)]{vf09} Vasudevan, R.V. \& Fabian, A.C. 2009, MNRAS, 392, 1124
\bibitem[Verner et al.(1996)]{vern} Verner, D.A., Ferland, G.J., Korista, K.T., \& Yakovlev, D.G. 1996, ApJ, 456, 487
\bibitem[Veilleux et al.(2009a)]{vei09a} Veilleux, S. et al. 2009a, ApJS, 182, 628 
\bibitem[Veilleux et al.(2009b)]{vei09b} Veilleux S. et al. 2009b, ApJ, 701, 587
\bibitem[Veilleux et al.(2013a)]{vei13a} Veilleux, S. et al. 2013a, ApJ, 764, 15
\bibitem[Weaver et al.(2010)]{weaver10} Weaver, K.A., Melendez, M., Mushotzky, R.F. et al. 2010, ApJ, 716, 1151
\bibitem[Weymann et al.(1991)]{weymann91} Weymann, R.J., Morris, S.L., Foltz, C.B., \& Hewett, P.C. 1991, ApJ, 373, 23
\bibitem[Wilms et al.(2000)]{wilm} Wilms, J., Allen, A., \& McCray, R. 2000, ApJ, 542, 914
\bibitem[Wu et al.(2010)]{wu10} Wu, J. et al. 2010, ApJ, 724, 762
\bibitem[Wu et al.(2011)]{wu11} Wu, J., Brandt, W.N., Hall, P.B., et al. 2011, ApJ, 736, 28
\bibitem[Yaqoob(2012)]{yaqoob12} Yaqoob, T. 2012, MNRAS, 423, 3360
\bibitem[Zdziarski (1988)]{zdziarski88} Zdziarski, A.A. 1988, MNRAS, 233, 739
\end{thebibliography}
\end{document}